\documentclass[hyper,12pt,a4paper]{JHEP3}
\pdfoutput=1

\usepackage{latexsym,amsmath,amsfonts,amssymb}
\usepackage{mathrsfs}
\usepackage{mathtools}
\numberwithin{equation}{section}

\newcommand{\be}{\begin{equation}} 
\newcommand{\ee}{\end{equation}}
\newcommand{\bea}{\begin{equation} \begin{aligned}} \newcommand{\eea}{\end{aligned} \end{equation}}

\newcommand{\bit}{\begin{itemize}} 
\newcommand{\eit}{\end{itemize}}

\newcommand{\cA}{\mathcal{A}}
\newcommand{\cB}{\mathcal{B}}

\newcommand{\cD}{\mathcal{D}}

\newcommand{\cG}{\mathcal{G}}
\newcommand{\cH}{\mathcal{H}}

\newcommand{\cJ}{\mathcal{J}}

\newcommand{\CL}{\mathcal{L}}

\newcommand{\cN}{\mathcal{N}}
\newcommand{\cO}{\mathcal{O}}
\newcommand{\cP}{\mathcal{P}}
\newcommand{\cQ}{\mathcal{Q}}
\newcommand{\cR}{\mathcal{R}}
\newcommand{\cS}{\mathcal{S}}

\newcommand{\cV}{\mathcal{V}}
\newcommand{\cW}{\mathcal{W}}

\newcommand{\cY}{\mathcal{Y}}
\newcommand{\cZ}{\mathcal{Z}}

\renewcommand{\t}{\widetilde }
\renewcommand{\d}{\partial }
\renewcommand{\b}{\bar}

\renewcommand\Re{\operatorname{Re}}
\renewcommand\Im{\operatorname{Im}}

\newcommand{\beq}{\begin{equation}}
\newcommand{\eeq}{\end{equation}}
\newcommand{\bal}{\begin{equation}\begin{aligned}}
\newcommand{\eal}{\end{aligned}\end{equation}}

\usepackage{color}

\usepackage{ulem}

\title{
The energy-momentum multiplet of supersymmetric defect field theories
}

\author{Nadav Drukker, Dario Martelli and Itamar Shamir
\\
Department of Mathematics, King's College London,\\
The Strand, WC2R 2LS, London, United Kingdom\\
\email{nadav.drukker@gmail.com},
\email{dario.martelli@kcl.ac.uk},
\email{itamar.shamir@kcl.ac.uk}}

\preprint{KCL-MTH-17-01}

\abstract{
Defects in field theories break translation invariance, 
resulting in the non-conservation of the energy-momentum tensor in the directions normal 
to the defect. This violation is known as the \emph{displacement operator}. 
We study 4d $\cN=1$ theories with 3d defects preserving 3d $\cN=1$ supersymmetry by analyzing the embedding of the 3d superspace in the 4d superspace. We use this to construct the energy-momentum multiplet of such defect field theories, 
which we call the \emph{defect multiplet} and show how it incorporates the displacement operator. We also derive the defect multiplet by using a superspace Noether procedure. 
}

\begin{document}

\maketitle

\tableofcontents



\section{Introduction}

In this paper we consider co-dimension one defects in 4d theories with $\cN =1$ supersymmetry. By this we mean 4d theories coupled to 3d  theories living on a 3d submanifold. We  focus on planar submanifolds, specified by a \textit{constant} and space-like normal vector $n^\mu$. The submanifold can be taken to be $x^n = x^\mu n_\mu=0$. The presence of the  defect leads to an explicit breaking of translation symmetry in the direction orthogonal to it. 
This manifests itself as a violation of the conservation of the energy-momentum tensor $T^{\mu\nu}$ by an operator local to the defect, which reads
\begin{align} \label{disp_def}
\d^\mu T_{\nu \mu} &= n_\nu \delta(x^n) f_d.
\end{align}
Here $f_d$ is called the displacement operator. The presence of the delta function means that away from the defect the energy-momentum tensor is conserved. 
Equation \eqref{disp_def} can be easily generalized to defects with co-dimension greater than one. We present two explicit examples of the displacement operator in bosonic theories in appendix~\ref{app:scalar}.

The displacement operator appears in several applications. The Bremsstrahlung function describing the radiation of an accelerating charge can be extracted as the coefficient of the two-point function of the displacement operator of a Wilson line \cite{Polyakov:2000ti,Semenoff:2004qr,Correa:2012at}. More recently, the displacement operator was used to study the dependence of entanglement entropy on the shape of the entangling surface \cite{Bianchi:2015liz,Dong:2016wcf,Balakrishnan:2016ttg,Bianchi:2016xvf,Chu:2016tps}. Additionally, 
conformal methods were used to constrain the form of correlation functions of the energy-momentum tensor and the displacement operator and to obtain constraints on the flow of defect field theories \cite{Jensen:2015swa,Billo:2016cpy,Gadde:2016fbj}. For co-dimension 2 defects in theories with $\cN=2$ supersymmetry in 4d, the displacement operator was discussed in \cite{Gaiotto:2013sma}.

Focusing on $\cN=1$ supersymmetry in 4d, the main goal of this paper is to construct the supersymmetric multiplet of the displacement operator. When there are no defects, it was shown that any $\cN=1$ theory in 4d admits a 
so-called $\cS$-multiplet \cite{Komargodski:2010rb}. It generalizes the Ferrara-Zumino (FZ), $\cR$ and superconformal multiplets which exist only under additional assumptions (see for instance \cite{Ferrara:1974pz,Gates:1983nr}). The $\cS$-multiplet may be  defined as a real vector superfield $\cS_{\alpha\dot\alpha}$ satisfying%
\footnote{Note that we use different conventions from \cite{Komargodski:2010rb}. In particular, bi-spinors are $\ell_{\alpha\dot\alpha} = \sigma^\mu_{\alpha\dot\alpha} \ell_\mu$, where we are using the notation of Wess and Bagger \cite{Wess:1992cp}.}
\begin{align} \label{S_mult}
\b D^{\dot \alpha} \cS_{\alpha\dot\alpha} = 2(\chi_\alpha - \cY_\alpha).
\end{align}
Here $\chi_\alpha$ is a chiral superfield satisfying $D^\alpha \chi_\alpha = \b D_{\dot\alpha} \b \chi^{\dot\alpha}$ and $\cY_\alpha$ is constrained by $\b D^2 \cY_\alpha =0$ and $D_{(\alpha} \cY_{\beta)}=0$. These conditions mean that we can locally solve $\cY_\alpha = D_\alpha X$ with $X$ chiral. 
An explicit computation shows that the components of $\cS_\mu$ include a symmetric and conserved energy-momentum tensor $T_{\nu\mu}$ and a conserved supercurrent $S_{\alpha\mu}$. Schematically, the component expansion of $\cS_\mu$ takes the form
\begin{align} \label{S_mult_exP_Int}
\cS_\mu = -i \theta (S_\mu + \ldots) + \theta \sigma^\nu \b \theta (2T_{\nu\mu} + \ldots) + \ldots
\end{align}

The main result of this paper is a modification of \eqref{S_mult} by terms arising from the presence of a defect. Since the defect necessarily breaks some of the translation and Lorentz symmetries it can at most preserve a subalgebra of supersymmetry. For $\cN=1$ in 4d and $n^\mu$ space-like the interesting cases are:
\begin{itemize}
\item co-dimension one defects preserving $\cN=1$ in 3d. 
\item co-dimension two defects preserving $\cN=(0,2)$ in 2d. 
\end{itemize}
Both these subalgebras preserve half of the original supersymmetries.
In this paper we consider the first case. 
We choose coordinates $x^\mu = (x^{i}, x^n)$ where $x^i$ are space-time coordinates, used along the world-volume of the defect. The preserved supercharges take the form $\hat Q_\alpha = \frac{1}{\sqrt{2}}\left(Q_\alpha + (\sigma^n \b Q)_\alpha \right)$ with
\begin{align} \label{N13d_subalg}
\{ \hat Q_\alpha, \hat Q_\beta \} = 2 (\Gamma^{ i})_{\alpha\beta} P_{i}.
\end{align}
Here the 3d gamma matrices are defined by $\Gamma^{i} \equiv \sigma^n \b \sigma^i $. 
Notice that only momenta orthogonal to $n^\mu$ appear in this algebra.

We propose the following modification of \eqref{S_mult}:
\begin{align} \label{def_mult}
\b D^{\dot \alpha} \cS_{\alpha\dot\alpha} = 2(\chi_\alpha - \cY_\alpha)+ 2\delta(\tilde y^n) \cZ_\alpha,
\end{align}
which we take as the definition of the \emph{defect multiplet}. Let us explain the ingredients which enter in the new term. The argument of the delta function is $\tilde y^n \equiv x^n + i \theta \sigma^n \b \theta - i \theta^2$. It has two virtues: (1) it is chiral 
(annihilated by $\b D_{\dot\alpha}$) and (2) it is invariant under the subalgebra \eqref{N13d_subalg}. This means that it breaks the symmetry in the correct way. We demand $\b D^2 \cZ_\alpha =0$ and the reality conditions
\begin{align} \label{}
\cZ_\alpha + (\sigma^n \b \cZ)_\alpha \xrightarrow[]{4 \to 3} 0,\qquad \quad 
\b D_{\dot\alpha} \cZ_\alpha + D_\alpha \b \cZ_{\dot\alpha} \xrightarrow[]{4 \to 3} -2i \sigma^n_{\alpha\dot\alpha} \mathscr{D}. 
\end{align}
The arrows imply a projection of the 4d superspace to the 3d $\cN=1$ superspace and $\mathscr{D}$ is a real scalar superfield of the 3d superspace. 

We show that \eqref{def_mult} implies the existence of an energy-momentum tensor satisfying \eqref{disp_def}
where $f_d$ is now the top component of $\mathscr{D}$. The energy-momentum tensor is conserved in the other directions, 
\emph{i.e.} $\d^{\mu} T_{j \mu}=0$, but it is generally not symmetric. Moreover, unlike the $\cS$-multiplet \eqref{S_mult_exP_Int} in which $S_{\alpha\mu}$ is a conserved supercurrent, in \eqref{def_mult} only the combination $S_{\alpha\mu} + \sigma_{\alpha\dot\alpha}^n \b S^{\dot\alpha}_\mu$ is conserved. This is the combination associated with the subalgebra \eqref{N13d_subalg}. 

In a purely 3d theory, the energy-momentum sits in a 3d $\cN=1$ multiplet analogous to the $\cS$-multiplet \eqref{S_mult}. Such multiplets were discussed in the literature in the superconformal case \cite{Kuzenko:2011xg,Kuzenko:2010rp,Buchbinder:2015qsa} (and for $\cN=2$ in 3d \cite{Dumitrescu:2011iu}). Using the 3d $\cN=1$ superspace coordinates $(x^i, \Theta_\alpha)$, where the Grassmannian coordinate is Majorana, satisfying the reality conditions $(\Theta_\alpha)^\dagger = \Theta^\alpha \sigma_{\alpha\dot\alpha}^n$, we define a 3d $\cN=1$ energy-momentum multiplet by
\begin{align} \label{3d_stress_mult}
\cD^\alpha \cJ_{\alpha j} = -2 \d_j \Sigma, \qquad
(\Gamma^j)_\alpha{}^\beta \cJ_{\beta j} = i \cD_\alpha( H - \Sigma),
\end{align}
where $\cD_\alpha$ is the covariant derivative in the 3d superspace. $\Sigma$ and $H$ are real scalar superfields and $\cJ_{\alpha j}$ is Majorana and for $\Sigma=H=0$ this multiplet reduces to the superconformal case. We show that \eqref{3d_stress_mult} leads to a component expansion, which includes
\begin{align} \label{}
\cJ_{\alpha j} = - S^{(3)}_{\alpha j} - i (\Gamma^i \Theta)_\alpha (2T^{(3)}_{ij} + \ldots) + \ldots,
\end{align}
where $S^{(3)}_{\alpha j}$ is a conserved Majorana supercurrent and $T^{(3)}_{ij}$ is a conserved and symmetric energy-momentum tensor. We study the structure of improvements of this multiplet and discuss two examples. 

When a defect field theory is constructed as a coupling of a 4d theory with a 3d theory, the total energy-momentum tensor of the system has a contribution localized on the defect
\begin{align} \label{T_total_int}
T_{\mu\nu} = T_{\mu\nu}^{(4)} + \delta(x^n) \cP_\mu{}^i \cP_\nu{}^j T_{ij}^{(3)},
\end{align}
where $\cP_n{}^i =0$ and $\cP_k{}^i = \delta_k{}^i$ is an embedding. The superspace analog of this statement, which is another result of this note, is that the 3d energy-momentum multiplet \eqref{3d_stress_mult} can be written as the $\cS$-multiplet in the 4d superspace. This is achieved by studying the embedding of the 3d superspace in the 4d one. We define a change of variables 
\begin{align} \label{}
\Theta_\alpha = \frac{1}{\sqrt2} (\theta + \sigma^n \b \theta)_\alpha, \qquad
\t\Theta_\alpha = \frac{i}{\sqrt2} ( \theta- \sigma^n \b \theta)_\alpha
\end{align}
in the 4d superspace and identify $\Theta_\alpha$ with the 3d Grassmannian coordinate. This allows us to embed \eqref{3d_stress_mult} in the $\cS$-multiplet as
\begin{align} \label{}
\cS^{(3)}_{\alpha\dot\alpha}  = \delta(\tilde x^n) \t\Theta^\beta \cJ_{\beta j} (\Gamma^j \sigma^n)_{\alpha\dot\alpha},
\end{align}
where $\tilde x^n = x^n - \Theta^\alpha \t \Theta_\alpha$ is an invariant of the subalgebra. This gives rise to the structure in \eqref{T_total_int}. When the 3d theory interacts with the 4d one, the $\cS$-multiplet must be modified to include the new term in \eqref{def_mult} which leads to the displacement operator. 

The outline of this paper is as follows. In section~\ref{sec:3d} we review the 3d $\cN=1$ superspace. In section~\ref{sec:EM3} we construct energy-momentum multiplets in 3d and discuss examples. In section~\ref{sec:34} we study the embedding of the 3d superspace in the 4d superspace as a tool for coupling 4d theories with 3d defect theories. We consider two representative examples: 4d chirals coupled to 3d scalars via a scalar potential and a bulk gauge multiplet coupled to a global symmetry on the defect. In section~\ref{sec:current} we consider global conserved currents as a simple application of the formalism developed. In section~\ref{sec:defect} we construct the defect multiplet. In section~\ref{sec:var} we show how to obtain the energy-momentum multiplets in 3 and 4 dimensions as well as the defect multiplet using a superspace Noether procedure. In section~\ref{sec:outlook} we discuss some applications and future directions. We include 3 appendices. In appendix~\ref{app:scalar} we review a computation of the displacement operator in two simple bosonic theories. Appendix~\ref{app:conv} includes two parts: In the first we review some necessary material on the 4d superspace, and in the second we collect some useful formulas corresponding to the embedding of the 3d superspace in 4d. Finally, in appendix~\ref{app:S} we review the $\cS$-multiplet as well as the example of chiral superfields which is used in the paper.


\section{$\cN=1$ supersymmetry in 3d}
\label{sec:3d}

In this section we review some basic facts about $\cN=1$ supersymmetry in 3d. Most of our presentation in this section is close in spirit to \cite{Gates:1983nr} although our conventions are different. The need to juggle two superspaces at the same time inevitably puts pressure on the available resources of letters and indices. We have chosen a minimalistic approach, whereby the reader is trusted with understanding from context which object lives in which universe. We hope this does not lead to much confusion. 

Let us begin by specifying our conventions for 3d, and their relation to 4d. It is important to emphasize that the constructions discussed in this section as well as the next one are strictly 3d. The invocation of the 4d embedding in our choice of conventions here is meant to facilitate the discussion of section~\ref{sec:34}, in which we consider the coupling of 3d and 4d theories. As described in the introduction, the embedding is specified by a constant \emph{space-like} vector $n_\mu$. This leads to a split $x^\mu = (x^n, x^i)$, where as before $x^n = n_\mu x^\mu$ and $x^i$ are coordinates of a 3d Minkowski space. Similarly, the 4d Pauli matrices split according to $\sigma^\mu_{\alpha\dot\alpha} = (\sigma_{\alpha\dot\alpha}^n, \sigma_{\alpha\dot\alpha}^i)$.%
\footnote{Our conventions for 4d are of course based on Wess and Bagger \cite{Wess:1992cp}. In particular we have the relation $\sigma^\mu \b \sigma^\nu = - \eta^{\mu\nu} + 2 \sigma^{\mu\nu}$, where $\sigma^{\mu\nu} = - \sigma^{\nu\mu}$. 
See appendix \ref{4d_superspace_app} for more useful 4d formulas.}
The basic spinor in 3d
is a Majorana doublet with a reality condition
\begin{align} \label{3d_reality}
\b \chi_{\dot\alpha} = (\chi_\alpha)^\dagger = (\chi \sigma^n)_{\dot\alpha}.
\end{align}
Even though, contrary to 4d, $\chi_\alpha$ and its conjugate transform in equivalent representations, it is convenient to keep track of dotted and undotted indices, which are converted by the use of $\sigma^n_{\alpha\dot\alpha}$.
In this way, spinor contraction as well as spinor indices lowering and raising follow straightforwardly from the 4d conventions. The 3d gamma matrices are $(\Gamma^{i})_\alpha{}^\beta = 2(\sigma^{n i})_\alpha{}^\beta$ and satisfy $ \Gamma^{i} \Gamma^{j} = -\eta^{ij} - i \epsilon^{i j k} \Gamma_{k}$. Here the 3d and 4d epsilon tensors are related by $\epsilon^{i j k}=\epsilon^{i j k n}$.%

The superspace coordinates are $(x^{i}, \Theta_\alpha)$ where $\Theta_\alpha$ are Grassmannian coordinates subject to the reality condition of eq. \eqref{3d_reality}. 
The supersymmetry generators in superspace are defined by
\begin{align} \label{3d_susy_gen}
\cQ_{\alpha} = \frac{\d}{\d \Theta^\alpha} - i(\Gamma^{j}\Theta)_\alpha \d_{j}, \qquad
\{ \cQ_\alpha, \cQ^\beta \} = 2i (\Gamma^{j})_\alpha{}^\beta \d_{j}.
\end{align}
We also define covariant derivatives by
\begin{align} \label{}
\cD_\alpha = \frac{\d}{\d \Theta^\alpha} + i(\Gamma^{j}\Theta)_\alpha \d_{j}, \qquad
\{ \cD_\alpha, \cD^\beta \} = -2i (\Gamma^{j})_\alpha{}^\beta \d_{j}.
\end{align}
As usual, the covariant derivatives are defined so that $\{ \cQ_\alpha, \cD_\beta \}=0$. Let us quote a few useful identities for the covariant derivatives
\begin{gather} \label{3d_cov_der_identity}
\cD_\alpha \cD_\beta = -i \d_{\alpha\beta} + \frac{1}{2}\epsilon_{\alpha\beta} \cD^2, \qquad 
\cD^\beta \cD_\alpha \cD_\beta =0, \\
\cD^2 \cD_\alpha = 2i \d_{\alpha}{}^\beta \cD_\beta, \qquad
\cD_\alpha \cD^2 = - 2i \d_{\alpha}{}^\beta \cD_\beta.
\end{gather}
Here the bi-spinor is defined as $\d_{\alpha}{}^{\beta} = (\Gamma^j)_{\alpha}{}^\beta \d_j$. We also use $(\Gamma^j)_{\alpha\beta}= \epsilon_{\beta\gamma} (\Gamma^{j})_\alpha{}^\gamma$, which is symmetric in the spinor indices. 

\subsection{Basic multiplets}

\subsubsection*{Scalar multiplet}
The simplest multiplet contains a real scalar, a Majorana fermion and a real auxiliary field. In superspace it is described by a scalar multiplet with the following component expansion
\beq \label{}
A = a + \Theta \chi + \frac{1}{2} \Theta^2 f_a.
\eeq
It is immediate to derive the supersymmetry variation $\delta A = \zeta \cQ A$ by using \eqref{3d_susy_gen}. We find
\bal \label{3dscalar_var}
&\delta a = \zeta \chi, \cr
&\delta \chi_\alpha = \zeta_\alpha f_a + i (\Gamma^{i} \zeta)_\alpha \d_{i} a, \cr
&\delta f_a = i \zeta \Gamma^{i} \d_{i} \chi. 
\eal

\subsubsection*{Vector multiplet}

Vector fields sit in a spinor multiplet $\cV_\alpha$ with gauge symmetry acting by $\delta \cV_\alpha= \cD_\alpha \omega$. The gauge symmetry can be used to fix the Wess-Zumino gauge, in which $\cV_\alpha$ takes the form
\begin{align} \label{3d_vec}
\cV_\alpha = i (\Gamma^{i} \Theta)_\alpha v_{i} - \Theta^2 \lambda_\alpha.
\end{align}
A gauge invariant field strength is defined by 
\begin{align} \label{3d_field_strength}
\cW_\alpha &= \frac{1}{2} \cD^\beta \cD_\alpha \cV_\beta = \lambda_\alpha - \frac{i}{2}\epsilon^{kij} (\Gamma_{k} \Theta)_\alpha F_{ij}
+ \frac{i}{2} \Theta^2 (\Gamma^{i} \d_{i} \lambda)_\alpha,
\end{align}
where $F_{ij}=\d_{i} v_{j} - \d_{j} v_{i}$. It follows immediately from the identity $\cD^\beta \cD_\alpha \cD_\beta=0$ that $\cD^\alpha \cW_\alpha=0$. In fact, this gives the Bianchi identity. 

Fully covariant derivatives are defined by $\mathscr{D}_\alpha = \cD_\alpha + i \cV_\alpha$ and $\mathscr{D}_{\alpha\beta} = \d_{\alpha\beta} + i \cV_{\alpha\beta}$, satisfying the following algebra
\begin{align} \label{cov_der_algebra}
\{ \mathscr{D}_\alpha, \mathscr{D}_\beta \} = -2i\mathscr{D}_{\alpha\beta}, \qquad
[ \mathscr{D}_\alpha, \mathscr{D}_{\beta\gamma} ] 
= - (\epsilon_{\alpha\beta}W_{\gamma} + \epsilon_{\alpha\gamma}W_\beta).
\end{align}
In particular, we have the relation $\cV_{\alpha\beta} = i\mathscr{D}_{(\alpha} \cV_{\beta)}$ with $\cV_{\alpha\beta}|=v_{\alpha\beta}$.

\subsubsection*{Current multiplet}

Conserved currents $\d_{i} j^{i} = 0$ sit in a spinor multiplet $J_\alpha$ satisfying $\cD^\alpha J_\alpha =0$. In components this is solved as
\begin{align} \label{3d_cons_current}
J_\alpha = \chi_\alpha + i (\Gamma^{i} \Theta)_\alpha j_{i} + \frac{i}{2} \Theta^2 (\Gamma^{i} \d_{i} \chi)_\alpha.
\end{align}
Clearly the field strength $\cW_\alpha$ defined above is a current multiplet, with the dual field strength conserved by the Bianchi identity. 

\subsection{Lagrangians and equations of motion}

A supersymmetric Lagrangian is a top component of a real scalar multiplet \eqref{3dscalar_var}. A simple example is $V_M = \frac{1}{2} \cD^\alpha A \cD_\alpha A$ which corresponds to a canonical kinetic term for $A$
\begin{align} \label{}
V_M|_{\Theta^2} &= - \frac{1}{2} (\d_j a)^2 - \frac{i}{2}\chi \Gamma^{j} \d_{j} \chi + \frac{1}{2} f^2_a.
\end{align}
This can be generalized to include multiple fields $A^I$ with a non-standard kinetic term $V_{M} = \frac{1}{2} \cG_{IJ}(A) \cD^\alpha A^I \cD_\alpha A^J$.  A scalar potential is constructed as a real function $P(A^I)$
and the equations of motion for such a model are
\begin{align} \label{3d_sgm_mdl_EOM}
\cD^2 A^I + \Gamma^I_{JK} \cD^\alpha A^J \cD_\alpha A^K = \cG^{IJ} \d_J P,
\end{align}
where $\Gamma^I_{JK}$ is the usual Levi-Civita connection of $\cG_{IJ}$.

A gauge invariant interaction can be succinctly written by promoting $\cD_\alpha$ to $\mathscr{D}_\alpha$ (see \eqref{cov_der_algebra} for the definition of $\mathscr{D}_\alpha$). For simplicity we shall stick to Abelian gauge fields and take $V_{MG} = \mathscr{D}^\alpha \b A \mathscr{D}_\alpha A$, where $A$ is here a complexified scalar multiplet.
The equation of motion is $\mathscr{D}^2 A=0$ and the gauge invariant current
\beq
\label{gauge-current}
J_{\alpha} = i(A\mathscr{D}_\alpha \b A - \b A \mathscr{D}_\alpha A)
\eeq
is obtained by the variation $\delta V_{MG} = \delta \cV^\alpha J_\alpha$. The Lagrangian for the gauge field is derived from the multiplet $V_G = -\frac{1}{2} \cW^\alpha \cW_\alpha$. More explicitly it is given by
\begin{align} \label{}
V_G|_{\Theta^2} = - \frac{i}{2} \lambda \Gamma^j \d_j \lambda - \frac{1}{4} F_{ij} F^{ij}.
\end{align}
The equation of motion is $\frac{1}{2}\cD^\beta \cD_{\alpha} \cW_\beta = J_\alpha$. It is also possible to include a Chern-Simons term $V_{CS} = \frac{\kappa}{2\pi}\cW^\alpha \cV_\alpha$, but being topological, it does not matter for anything we do in the sequel.

\section{Energy-momentum multiplets in 3d}
\label{sec:EM3}

Any local supersymmetric field theory contains a conserved and symmetric energy-momentum tensor $T^{(3)}_{ij}$ and a conserved supercurrent $S^{(3)}_{\alpha i}$. It follows from the algebra of supersymmetry that these two operators sit in the same multiplet. When a superspace realization of the algebra is available, then these operators can be incorporated in a superfield. In this section we define such superfields. All 3d $\cN=1$ theories admit a maximal multiplet with $6+6$ components, but some theories admit shorter multiplets with $4+4$ or $2+2$ components (in the superconformal case). The superconformal multiplet was described by several groups before \cite{Kuzenko:2011xg,Kuzenko:2010rp,Buchbinder:2015qsa}, which also discussed extended supersymmetry. However, to the best of our knowledge the 
non-conformal energy-momentum multiplets were not considered elsewhere in the literature. (See \cite{Dumitrescu:2011iu} for 3d $\cN=2$.)
We discuss the structure of improvements of these multiplets and review some examples. 

We define a real multiplet $\cJ_{\alpha i}$ by
\begin{equation} \label{3d_S_mult}
\cD^\alpha \cJ_{\alpha i} = -2\d_{i} \Sigma, \qquad
(\Gamma^{i})_\alpha{}^\beta \cJ_{\beta i}= i \cD_\alpha (H-\Sigma),
\end{equation}
with $\Sigma$ and $H$ both real multiplets. The component expansion of $\Sigma$ and $H$ is
\begin{equation} \label{}
\Sigma = \sigma + \Theta \psi + \frac{1}{2}\Theta^2 f_\sigma, \qquad
H =  \eta + \Theta \kappa + \frac{1}{2}\Theta^2 f_\eta.
\end{equation}
Let us emphasize that only the derivatives of $\sigma$ and $\eta$ are guaranteed to be well-defined. More generally we can write $\cD^\alpha \cJ_{\alpha i} =-2 \Sigma_{i}$ and $\cJ_{\beta\alpha}{}^{\beta} = i \cH_\alpha$ which we require to satisfy $\d_{[i} \Sigma_{j]}=0$ and $\cD^\alpha \cD_\beta \cH_\alpha=0$. This means that locally we can solve $\Sigma_{i} = \d_{i}\Sigma$ and $\cH_\alpha = \cD_\alpha (H - \Sigma)$. To simplify the notation we will not make this explicit. 

Solving \eqref{3d_S_mult} for the components of $\cJ_{\alpha i}$ we obtain 
\bal \label{3d_S_mult_exp}
\cJ_{\alpha j} &= - S^{(3)}_{\alpha j} + i (\Gamma_{j} \psi)_\alpha + \Theta_\alpha \d_{j} \sigma 
- i (\Gamma^{i} \Theta)_\alpha \left( 2T^{(3)}_{ij} - \eta_{ij}f_\sigma + \frac{1}{2} \epsilon_{ijk} \d^{k}\eta\right) \cr
&\quad - \frac{1}{2} \Theta^2 \left( i (\Gamma^{i} \d_{i} S_{j})_\alpha - (\Gamma_{j} \Gamma^{i} \d_{i} \psi)_\alpha \right).
\eal
Here $T^{(3)}_{ij}$ is symmetric and conserved, $S^{(3)}_{\alpha i}$ is conserved, and the following relations hold
\beq \label{}
T^{(3)i}{}_{i} = f_\eta + 2 f_\sigma, \qquad
S^{(3)}_{\alpha \beta}{}^\alpha = i\left(\kappa_\beta +2 \psi_\beta \right).
\eeq
The combination $H+ 2 \Sigma$ is the `trace multiplet'. Before continuing, let us mention that a simple generalization of \eqref{3d_S_mult} is obtained by changing the second equation to $\cJ_{\beta\alpha}{}^\beta = i \cD_{\alpha}(H - \Sigma) + i J_\alpha$, where $J_\alpha$ is a conserved current. This is a multiplet which encompasses a non-symmetric energy-momentum tensor.

\subsection{Improvements}

Let us examine how this multiplet can be modified. For any real multiplet $U=u + \Theta \rho + \frac{1}{2} \Theta^2 f_u$, we can act on $\cJ_{\alpha i}$ by the following transformation
\begin{align} \label{3d_imp1}
\delta \cJ_{\alpha i} = i (\Gamma_{i})_{\alpha}{}^\beta \cD_\beta U, \qquad
\delta \Sigma =  U, \qquad
\delta H = -2  U,
\end{align}
under which the energy-momentum tensor and the supercurrent do not change. In particular, it is easy to see that the `trace multiplet' remains unmodified. Another way to transform the multiplet is by
\begin{align} \label{3d_imp2}
\delta \cJ_{\alpha i} = -2\d_{i} \cD_\alpha U, \qquad 
\delta \Sigma = \cD^2 U, \qquad
\delta H = 0,
\end{align}
which is a \emph{bona fide} improvement. The resulting transformation is
\begin{align} \label{3d_imp3}
\delta S^{(3)}_{\alpha}{}^k = 2 (\Gamma^{[k} \Gamma^{j]} \d_{j} \rho)_\alpha 
, \qquad
\delta T^{(3)}_{ij} = ( \d_{i} \d_{j} - \eta_{ij} \d^2)u.
\end{align}

The general $\cJ_{\alpha i}$ multiplet we obtained, has $6+6$ components, but is not minimal (similarly to the $\cS$-multiplet is 4d). A submultiplet with $4+4$ components can be achieved by using equation \eqref{3d_imp1} to set some linear relation between $\Sigma$ and $H$. The form of the improvement \eqref{3d_imp2} suggests that a natural choice is $H=0$. This is achieved by taking $U=\frac{1}{2}H$ in \eqref{3d_imp1}, which results in the multiplet
\begin{align} \label{3d_mult_short}
\cD^\alpha \cJ_{\alpha i} = -2\d_{i} \Sigma, \qquad
(\Gamma^{i})_\alpha{}^\beta \cJ_{\beta i}= -i \cD_\alpha \Sigma.
\end{align}
As explained by Komargodski and Seiberg \cite{Komargodski:2010rb} such improvements only make sense if $H$ is a well-defined operator (\emph{e.g.} gauge invariant). It is interesting that in the examples we study below, $H$ indeed turns out to be well-defined, perhaps suggesting the \eqref{3d_mult_short} always exists. We will see in section \ref{sec:defect} that this multiplet is closely related to the 4d FZ multiplet.

In some theories we may be able to further shorten the multiplet using the improvement \eqref{3d_imp2}. This is possible if and only if there is a well-defined $U$ such that $\Sigma=\cD^2 U$. When this is the case, we can set to zero the `trace multiplet' leading to a multiplet satisfying $\cD^\alpha \cJ_{\alpha i} =0$ and $(\Gamma^{i})_\alpha{}^\beta \cJ_{\beta i}=0$ which is a $2+2$ superconformal multiplet (see \cite{Kuzenko:2011xg,Kuzenko:2010rp,Buchbinder:2015qsa}).

\subsection{Examples}
As a first example we consider the sigma model of scalar multiplets $A^I$ described in the previous section. 
The energy-momentum multiplet is then given by 
\begin{align} \label{sigma_model_mult}
\cJ_{\alpha i} = -2 \cG_{IJ} \d_{i} A^I \cD_\alpha A^J, \qquad
\Sigma= V_M+ P, \qquad
H = -V_M.
\end{align}
The bottom component of $P$ is not necessarily a well-defined operator. (For example, as in \cite{Dumitrescu:2011iu} we can take $A^I \sim A^I + 1$ and a linear potential.) However, we note that $H$ is well-defined, and we can define the shorter multiplet \eqref{3d_mult_short}, in which $\Sigma=\frac{1}{2}V_M+P$. If the theory is free and massless, \emph{i.e.} $G_{IJ} = \delta_{IJ}$ and $P=0$, we can use the equations of motion to write $\Sigma = \frac{1}{4}\cD^2 (A^IA_I)$ which means that the theory admits a superconformal multiplet. 

Next, consider an Abelian gauge field coupled to a complex scalar multiplet. The matter and gauge contributions to the energy-momentum multiplet are given by
\begin{align} 
(\cJ_{MG})_{\alpha\beta\gamma} &= - 2(\mathscr{D}_{\beta\gamma} \b A \mathscr{D}_\alpha A + \mathscr{D}_{\beta\gamma} A \mathscr{D}_\alpha \b A), \label{3d_MG_mult1} \\
(\cJ_G)_{\alpha\beta\gamma} &= - i \cW_\alpha \cD_{(\beta} \cW_{\gamma)}, \label{3d_MG_mult2}
\end{align}
and satisfy
\begin{align} 
&\cD^\alpha (\cJ_{MG})_{\alpha\beta\gamma} = -2\d_{\beta\gamma} (\mathscr{D}^\alpha \b A \mathscr{D}_\alpha A) - 4i \cW_{(\beta} J_{\gamma)}, \label{3d_MG_mult3} \\
&\cD^\alpha (\cJ_G)_{\alpha\beta\gamma} = -2\d_{\beta\gamma} \left(\frac{1}{2} \cW^\alpha \cW_\alpha \right) + 4i \cW_{(\beta} J_{\gamma)}, \label{3d_MG_mult4}
\\
&\label{3d_MG_mult5}
(\cJ_{MG})_{\alpha\beta}{}^\alpha = -2i D_\beta(\mathscr{D}^\alpha \b A \mathscr{D}_\alpha A),
\qquad
(\cJ_G)_{\alpha\beta}{}^\alpha = -i D_\beta(\cW^\alpha \cW_\alpha),
\end{align}
where $J_\alpha$ is the gauge current \eqref{gauge-current}. We can identify that $\Sigma_{MG} = - H_{MG} = V_{MG}$ and $\Sigma_G = -H_G = - V_G$, where $V_{MG}$ and $V_G$ are the matter and gauge kinetic terms respectively and are defined in the previous section.

Let us note the cross-term $4i \cW_{(\beta} J_{\gamma)}$ in \eqref{3d_MG_mult3}, which clearly cancels in the sum with \eqref{3d_MG_mult4}. We have separated here the contributions of the matter part and the gauge part deliberately to exhibit this term. 
The reason is that, unlike the case here, when we couple the 3d matter fields to 4d gauge fields below, 
then the term in \eqref{3d_MG_mult3} will not be sufficient to completely cancel the 4d contribution. The remainder will be identified as the displacement operator.


\section{Coupling of 3d and 4d theories}
\label{sec:34}

In the previous two sections we have discussed various aspects of theories with $\cN=1$ supersymmetry in 3d. We are now ready to begin our exploration of the main theme of this paper: the supersymmetric coupling of 3d and 4d theories, and the structure of their combined energy-momentum multiplet. In this section we will explain how this supersymmetric coupling can be performed in a manifest fashion, via superspace.

Some of the discussion parallels a previous work by Bilal \cite{Bilal:2011gp}, who considered 4d $\cN=1$ theories with a boundary preserving a 3d $\cN=1$ subalgebra (see also \cite{DeWolfe:2001pq,Mintun:2014aka} and \cite{Erdmenger:2002ex,Constable:2002xt} in 4d $\cN=2$). But there are also important differences, to be pointed out below, which are crucial to the main goals of this paper. For example, we demonstrate how to write superspace equations of motion for the coupled system.

The basis for constructing supersymmetric coupling of 3d and 4d theories is to study the 3d superspace embedding in the 4d superspace. A simple approach utilizes the pattern of symmetry breaking. As discussed in the introduction, the embedding can preserve at most two supersymmetries. Clearly, the broken Poincar\'e symmetries can be used to fix the normal vector $n^\mu$ to some specified direction and translate the 3d subspace to a point in the $x^n$ direction, say the origin. Supersymmetry acts on the 4d superspace coordinates $(x^\mu, \theta_\alpha, \b \theta_{\dot\alpha})$ by
\begin{align} \label{action_superspace}
\delta x^\mu = i\theta \sigma^\mu \b \zeta - i \zeta \sigma^\mu \b\theta, \qquad
\delta \theta_\alpha = \zeta_\alpha, \qquad
\delta \b \theta_{\dot\alpha} = \b \zeta_{\dot\alpha}. 
\end{align}
The subalgebra we consider is determined by the relation $\zeta_\alpha = (\sigma^n \b \zeta)_\alpha$.%
\footnote{The broken $R$-symmetry corresponds to a possible phase $\zeta_\alpha = e^{i\eta} (\sigma^n \b \zeta)_\alpha$. If the 4d field theory that we consider has an $R$-symmetry, we can use it to dial $\eta=0$. Otherwise, it is a genuine parameter of the embedding. In any event, we shall keep using $\eta=0$ to simplify the notation.}
Equivalently, we consider a supersymmetry generator which is a linear combination of supercharges with opposite chirality
\begin{align} \label{}
\hat Q_\alpha = \frac{1}{\sqrt2} \left(Q_\alpha + (\sigma^n \b Q)_\alpha \right)
\end{align}
As can be seen from \eqref{N13d_subalg} this generates an algebra isomorphic to $\cN=1$ in 3d. The action leaves the following combinations of superspace coordinates invariant 
\begin{align} \label{}
\tilde x^n \equiv x^n - \frac{i}{2}(\theta^2 - \b \theta^2), \qquad
\t \Theta_\alpha = \frac{i}{\sqrt{2}}(\theta - \sigma^n \b \theta)_\alpha. 
\end{align}
We can use these two coordinates to generate other invariants. For example, 
\begin{align} \label{}
\tilde y^n \equiv \tilde x^n + i \t \Theta^2
\end{align}
is a chiral combination. It is illuminating to write $\tilde y^n$ in terms of the chiral coordinate of superspace $y^\mu = x^\mu + i\theta\sigma^\mu \b \theta$. We find $\tilde y^n = y^n - i \theta^2$, which is clearly a chiral combination.%
\footnote{For example, a chiral superfield $\Phi=(\phi,\psi_\alpha,F)$ invariant under $\hat Q_\alpha$ must be a function of $\tilde y^n$, leading to the component expansion $\Phi(\tilde y^n) = \phi(y^n) - i \theta^2 \d_n \phi(y^n)$.
This is explained as follows. Observing the variations of $\Phi$ and demanding invariance 
implies $\psi_\alpha=0$ and setting
$$
\delta \psi_\alpha 
= \sqrt{2} \left[ \zeta_\alpha (F+ i \d_n \phi) +2i (\sigma^{n\mu} \zeta)_\alpha \d_\mu \phi \right]
$$
to zero means that $\phi=\phi(x^n)$
and $F= - i \d_n \phi$. This is precisely the component expansion above.%
} 
We can complete $\tilde x^n$ and $\t \Theta_\alpha$ to a basis of the 4d superspace by including $x^{i}$ and another Grassmann coordinate 
\begin{align} \label{}
\Theta_\alpha = \frac{1}{\sqrt{2}}(\theta + \sigma^n \b \theta)_\alpha. 
\end{align}
In total, we have the change of basis 
\begin{align} \label{}
(x^\mu, \theta_\alpha , \b \theta_{\dot\alpha}) \longleftrightarrow (x^{i}, \tilde x^n, \Theta_\alpha, \t \Theta_\alpha).
\end{align}
These coordinates are natural from the point of view of the preserved subalgebra. Clearly, $(x^{i}, \Theta_\alpha)$ can be identified with the coordinates in the 3d superspace, and are acted upon by the subalgebra in the expected way. Unlike \cite{Bilal:2011gp,DeWolfe:2001pq} where the 3d superspace is identified via the relation $\theta_\alpha = (\sigma^n \b \theta)_\alpha$ (or simply $\t \Theta_\alpha=0$ in our language), we here find it very useful to keep track of all the coordinates including $\t\Theta_\alpha$. It will become apparent below why this is advantageous.

Consider now a general superfield $F(x^\mu,\theta, \b \theta)$. We would like to understand how to decompose it into representations of the subalgebra. This is obtained by writing the superfield in the coordinates system introduced above and expanding in $\t\Theta_\alpha$
\begin{align} \label{comp_superfield}
F(x^i, \tilde x^n, \Theta, \t\Theta) = F_{1}(x^i, \tilde x^n, \Theta) + \t\Theta^\alpha F_{2\alpha}(x^i, \tilde x^n, \Theta) + \frac{1}{2} \t\Theta^2 F_{3}(x^i, \tilde x^n, \Theta).
\end{align}
It is obvious from the discussion above that the component superfields transform independently under the subalgebra. However, for practical reasons it is usually more convenient to work with fields which are functions of $x^n$ instead of $\tilde x^n$ (note that $\tilde x^n = x^n - \t\Theta \Theta$). Namely, in the coordinate system $(x^{i}, x^n, \Theta_\alpha, \t \Theta_\alpha)$. However, this brings about a small complication, as one observes by writing explicitly the preserved supercharge in these coordinates $\hat Q_\alpha = \cQ_\alpha + \t \Theta_\alpha \d_n$. Here $\cQ_\alpha$ is the 3d expression in \eqref{3d_susy_gen}. In other words, this means that component superfields in a $\t\Theta_\alpha$ expansion mix under $\hat Q_\alpha$. As usual, the problem is solved by introducing covariant derivatives. We define
\bal\label{new_cov_der}
\Delta_\alpha &\equiv \frac{1}{\sqrt{2}} \left(D_\alpha + (\sigma^n \b D)_\alpha \right) 
= \frac{\d}{\d \Theta^\alpha} + i(\Gamma^{i}\Theta)_\alpha \d_{i} - \t \Theta_\alpha \d_n, \cr
\t \Delta_\alpha &\equiv -\frac{i}{\sqrt{2}} \left(D_\alpha - (\sigma^n \b D)_\alpha \right)
= \frac{\d}{\d \t \Theta^\alpha} + i(\Gamma^{i}\t \Theta)_\alpha \d_{i} + \Theta_\alpha \d_n.
\eal
By construction we have that $\{\hat Q_\alpha, \Delta_\alpha\} = \{\hat Q_\alpha, \t\Delta_\alpha\}=0$. The component superfields of \eqref{comp_superfield} can thus be obtained by taking $\t \Delta_\alpha$ derivatives and projecting to the 3d superspace by setting $\t\Theta=0$. Strictly speaking, a projection should also include $x^n=0$ (or some other point). Nevertheless, it is convenient to keep the location of the defect unspecified, \emph{i.e.} keep explicit dependence on $x^n$.

The simplest example is that of a chiral superfield $\Phi=(\phi,\psi,F)$, for which we obtain%
\footnote{Let us note the relations
$$
\theta \sigma^n \b \theta = \frac{1}{2} (\Theta^2 +\t \Theta^2), \qquad
\theta \sigma^i \b \theta =  - i\t \Theta \Gamma^{i} \Theta.
$$
We refer the reader to appendix \ref{3d_superspace_app} for more superspace relations.}
\bal \label{43chiral_proj}
\Phi(y^{i}, y^n,\theta) |_{\t \Theta=0} =
\Phi (x^{i}, x^n+ \tfrac{i}{2}\Theta^2, \tfrac{1}{\sqrt{2}}\Theta) 
=\phi + \Theta \psi + \frac{1}{2} \Theta^2(F+ i\d_n \phi). 
\eal
A similar expression for this projection was obtained in \cite{Bilal:2011gp,DeWolfe:2001pq}. For a chiral superfield $\t\Delta_\alpha$ does not give a new superfield since from \eqref{new_cov_der} it has the same effect as $\Delta_\alpha$.
Similarly, the anti-chiral superfield projects to
\bal\label{43a_chiral_proj}
\b\Phi(\b y^{i},\b y^n,\b \theta) |_{\t \Theta =0} = 
\b \Phi (x^{i}, x^n -\tfrac{i}{2}\Theta^2, \tfrac{1}{\sqrt{2}}\Theta\sigma^n) 
=\b \phi + \Theta \sigma^n \b\psi + \frac{1}{2} \Theta^2(\b F- i\d_n \b\phi).
\eal
Conversely, given a 3d superfield (with or without $x^n$ dependence) we can embed it 
into the 4d superspace. As demonstrated below, this is required in order to write equations of motion for the coupled system in the 4d superspace. As a simple example, a 3d scalar multiplet $A=(a,\chi,f_a)$ can be embedded as a chiral multiplet by
\bal
A(x^{i},x^n, \Theta) \xrightarrow[]{3 \to 4} \cA(y,\theta) &\equiv A (y^{i},\tilde y^n,\sqrt{2}\theta) \\
&= a + \sqrt{2}\,\theta \chi + \theta^2 (f_a - i \d_n a).
\eal
Let us remark that this is a ``real chiral superfield''. Its existence is a by-product of the coupling to 3d and will be important in the sequel. 
It is useful for later computations to show more explicitly the relation between $A$ and $\cA$. This is achieved by expanding around $(x^{i},\Theta)$ in the following way
\bal \label{chiral_emb_exp}
\cA = A(y,\sqrt{2}\theta) 
&= A(x^{i} + \t\Theta \Gamma^{i} \Theta, \tilde x^n + i \t \Theta^2, \Theta - i \t \Theta) \cr
&= A(x^{i},\tilde x^n, \Theta) - i \t \Theta^\alpha \Delta_\alpha A
+ \frac{1}{4} \t\Theta^2\Delta^2 A
\eal
This relation shows that $\cA$ is the unique chiral superfield whose projection $\t\Theta=0$ is $A$.
It is also useful as a trick to simplify certain computations below. In a similar way, we can embed $A$ in an anti-chiral superfield $\b\cA \equiv A(\b y^{i},\tilde{\b y}^n,\sqrt{2}\sigma^n\b\theta)$.

We note that projecting a chiral (anti-chiral) superfield to 3d and then lifting it to a chiral (anti-chiral) returns the original field. However, if we start with an anti-chiral $\b \Phi$, project to 3d $\b \Phi |_{\t\Theta=0}$ and then lift to a \emph{chiral} we get
\begin{align} \label{chiral_anti_chiral}
\t{\b \Phi}(y, \theta) = \b \phi + \sqrt{2} \theta \sigma^n \b \psi + \theta^2 (\b F - 2 i \d_n \b \phi).
\end{align}
It is easy to check that under the subalgebra \eqref{N13d_subalg} the multiplet $\t{\b \Phi} = (\b \phi, \sigma^n \b \psi, \b F - 2i \d_n \b \phi)$ transforms as a chiral. In the same sense $\t \Phi = (\phi, - \b \sigma^n \psi, F+ 2i\d_n \phi)$ is an anti-chiral superfield.

The embedding of a 3d superfield in 4d superspace can be written in another way, that 
is more useful for computations. 
Starting with a superfield $A(x^{i},x^n, \Theta)$ we define 4d chiral and anti-chiral superfields by
\begin{align} \label{chi_emb_formula}
A \xrightarrow[]{3 \to 4} \cA \equiv \frac{1}{2} \b D^2( \t \Theta^2 A), \qquad
A \xrightarrow[]{3 \to 4} \b \cA \equiv \frac{1}{2} D^2( \t \Theta^2 A).
\end{align}
Clearly $\cA$ and $\b \cA$ are chiral and anti-chiral superfield respectively. With some labour this can be computed explicitly and shown to be equivalent to the expansion \eqref{chiral_emb_exp} (and similarly for the anti-chiral). However, a simple trick renders this computation trivial. 
Because of the $\t \Theta^2$ factor we can change the arguments of $A$ 
in the chiral embedding by $\t\Theta_\alpha$ terms without changing the expression. There is a unique way of doing it which makes $A$ chiral, namely $A(y^{i}, \tilde y^n, \sqrt{2}\theta)$. Then $\b D^2$ acts only on $\t \Theta^2$ and the result follows. 

Another useful relation allows us to rewrite 3d Lagrangians in 4d superspace. Recall that a 3d Lagrangian is a top component of a real scalar multiplet. Multiplying by $\t \Theta^2$ allows us to write this as a $D$-term of a 4d real multiplet. Specifically, let $P = p + \Theta \chi+ \frac{1}{2} \Theta^2 f_p$, then $(-1)\t\Theta^2 P = \ldots + \frac{1}{2} \theta^2 \b \theta^2 f_p$. We therefore have the prescription
\begin{align} \label{}
\int d^3 x \int d^2 \Theta P = \int d^4 x \int d^4 \theta (-1)\delta(\tilde x^n)\t\Theta^2 P.
\end{align}
Notice that $\t\Theta^2$ can be thought of as a Grassmannian delta function. More generally, we can replace $- \delta(\tilde x^n)\t\Theta^2$ by any function $f = f(\tilde x^n, \t\Theta)$ without breaking the symmetry further. This can be interpreted as a smeared defect.

To do the same for gauge fields, consider first a real multiplet $V$ with components $(C,\chi,M,v_\mu,\lambda,D)$
which we decompose following the procedure given above. Most interesting is the component containing the vector. It is given by
\bal\label{}
\t\Delta_\alpha V|_{\t\Theta=0} &= \frac{1}{\sqrt2}(\chi + \sigma^n \b \chi)_\alpha + \Theta_\alpha \left(\frac{1}{2}(M+\b M) + \d_n C \right) + i (\Gamma^{i} \Theta)_\alpha v_{i}  \\
&\quad - \frac{1}{2}\Theta^2 \left(\sqrt2( \lambda + \sigma^n \b\lambda) + \frac{i}{\sqrt2}\Gamma^{i} \d_{i}(\chi+ \sigma^n \b \chi) \right).
\eal
This multiplet can be identified with the 3d vector multiplet $\cV_\alpha$ mentioned in the previous sections. In particular the gauge symmetry of the real multiplet $\delta V = \frac{i}{2}(\Omega- \b \Omega)$ translates into
\begin{align} \label{}
\t\Delta_\alpha \delta V|_{\t\Theta=0} &= \frac{1}{2\sqrt{2}}(D - \sigma^n \b D)_\alpha (\Omega - \b \Omega)|_{\t\Theta=0} 
= \frac{1}{2}\Delta_\alpha (\Omega+ \b \Omega)|_{\t\Theta=0},
\end{align}
where $\Omega$ is a chiral superfield.
We can identify the 3d gauge parameter multiplet $\omega$ with the real scalar multiplet $\omega= \frac{1}{2}(\Omega + \b \Omega)|_{\t\Theta=0}$ as above equation \eqref{3d_vec}. 

Next consider the field strength $W_\alpha = - \frac{1}{4} \b D^2 D_\alpha V$ which satisfies by construction $D^\alpha W_\alpha = \b D_{\dot\alpha} \b W^{\dot\alpha}$. It is decomposed as
\begin{align} \label{}
\cW_\alpha \equiv \frac{i}{\sqrt2}(W-\sigma^n \b W)_\alpha|_{\t\Theta=0}, \qquad 
\t\cW_\alpha \equiv \frac{1}{\sqrt2}(W+\sigma^n \b W)_\alpha|_{\t\Theta=0}.
\end{align}
Expanding in components they give
\begin{align} \label{}
\cW_\alpha &= \frac{1}{\sqrt2}(\lambda + \sigma^n \b \lambda)_\alpha - \frac{i}{2} \epsilon^{kijn} (\Gamma_{k} \Theta)_\alpha F_{ij} + \frac{i}{2\sqrt2} \Theta^2 \Gamma^{i} \d_{i} (\lambda+\sigma^n \b\lambda)_\alpha, \\
\t\cW_\alpha &= -\frac{i}{\sqrt2}(\lambda - \sigma^n \b \lambda)_\alpha + \Theta_\alpha D - i (\Gamma^{i}\Theta)_\alpha F_{n i} \cr
&\quad \qquad \qquad + \frac{1}{2} \Theta^2 \left(- \frac{1}{\sqrt2}\Gamma^{i} \d_{i}(\lambda - \sigma^n \b\lambda)_\alpha
+ \sqrt{2} \d_n (\lambda + \sigma^n \b \lambda)_\alpha\right).
\end{align}
Clearly $\cW_\alpha$ can be identified with the field strength defined in 3d \eqref{3d_field_strength}. In particular we have $\Delta^\alpha \cW_\alpha = \frac{i}{2}(D^\alpha W_\alpha - \b D_{\dot\alpha} \b W^{\dot\alpha})=0$.

\subsection{Example 1 -- scalar multiplets}

Consider 4d chiral superfields $\Phi^a$ with a K\"{a}hler potential $K(\Phi^a,\b \Phi^{\b a})$ and a superpotential $W(\Phi^a)$.
On the defect we consider real scalars $A^I$ with a kinetic term $V_M = \frac{1}{2} \cG_{ij} \Delta^\alpha A^I \Delta_\alpha A^J$ as described around \eqref{3d_sgm_mdl_EOM}.%
\footnote{Note that since $A^I$ are 3d fields (independent of $x^n$), $\cD_\alpha$ and $\Delta_\alpha$ can be used interchangeably.}
The 3d and the 4d theories interact through a potential $P(\Phi^a,\b \Phi^{\b a}, A^I)|_{\t\Theta=0}$ localized on the defect.%
\footnote{A similar potential was considered in \cite{Bilal:2011gp} as a boundary interaction.}
It is clear that the 3d equations of motion stay the same as in \eqref{3d_sgm_mdl_EOM} with $P(\Phi^a,\b \Phi^{\b a}, A^I)|_{\t\Theta=0}$ substituting for the purely 3d potential. We would also like to obtain the equations of motion of $\Phi^a$ with the defect interaction. Following the discussion above we can lift the potential to the 4d superspace by
\begin{align} \label{}
\int d^3x \int d^2 \Theta P|_{\t\Theta=0} = \int d^4x \int d^2 \theta d^2 \b \theta \left((-1)\delta(\tilde x^n) \t \Theta^2 P \right).
\end{align}
Here we switched $\tilde x^n$ for $x^n$ in the delta function so that it manifestly preserves the desired symmetries. 
The difference is proportional to $\t \Theta_\alpha$ and does not change the expression. To compute the equations of motion we change to integration over half superspace
\begin{align} \label{}
\int d^2 \theta d^2 \b \theta \left((-1)\delta(\tilde x^n) \t \Theta^2 P \right) = \int d^2 \theta \b D^2 \left( \frac{1}{4}\delta(\tilde x^n) \t \Theta^2 P \right),
\end{align}
and use the relation
\begin{align} \label{chiral_lift}
\frac{1}{2} \b D^2 \left( \delta(\tilde x^n) \t \Theta^2 P(\Phi, \b \Phi, A) \right) = \delta(\tilde y^n) \cP(\Phi, \t{\b \Phi},\cA).
\end{align}
Here $\cA$ and $\cP$ are the chiral lifts of $A$ and $P$ and $\t{\b \Phi}$ is the chiral associated with the anti-chiral $\b \Phi$ as per \eqref{chiral_anti_chiral}. This leads to the equation of motion
\begin{align} \label{4d_pot_EOM}
\b D^2 K_a = 4 W_a + 2\delta(\tilde y^n) \cP_a.
\end{align}
Clearly for this equation to make sense the delta function must be a chiral superfield.

\subsection{Example 2 -- gauge interactions}

Consider a 4d $U(1)$ gauge theory. As demonstrated above, $\cV_\alpha \equiv \t\Delta_\alpha V|_{\t\Theta=0}$ is equivalent to a 3d gauge multiplet. Therefore we can take a 3d theory with a global $U(1)$ symmetry and gauge it by coupling to the $U(1)$ gauge field coming from 4d. 
The coupling is identical to the minimal coupling for a complexified scalar multiplet $A$ considered above \eqref{gauge-current} and so are the resulting 3d equations of motion for $A$. Here we obtain the 4d gauge field equation of motion coupled to the 3d matter current \eqref{gauge-current}. 

As usual the 4d gauge part is given by $\frac{1}{4} \int d^2 \theta W W + \frac{1}{4} \int d^2 \b \theta \b W \b W$. The unconstrained variable which we must vary to obtain the equation of motion is $V$. By standard superspace maneuvers we obtain
\begin{align} \label{}
\int d^4 \theta \,\delta V \left( - \frac{1}{2} (D^\alpha W_\alpha + \b D_{\dot\alpha} \b W^{\dot\alpha} ) \right).
\end{align}
For comparison, consider a charged 4d chiral field $\Phi$. The Lagrangian is $\int d^4 \theta \b \Phi e^{2V}\Phi$. Identifying the 4d current as $J = \b \Phi e^{2V}\Phi$, the contribution to the equations of motion is
\begin{align} \label{}
- \frac{1}{2} (D^\alpha W_\alpha + \b D_{\dot\alpha} \b W^{\dot\alpha}) = 2 J.
\end{align}
Similarly, from \eqref{gauge-current}, the variation of the 3d coupling gives $\int d^2 \Theta \delta \cV^\alpha J_\alpha$, which upon lifting to 4d becomes 
\begin{align} \label{}
-\int d^4 \theta \delta(\tilde x^n) \t\Theta^2 \delta \cV^\alpha J_\alpha
= -\int d^4 \theta \delta V 2\delta(\tilde x^n)\t\Theta^\alpha J_\alpha.
\end{align}
Here we have used the relation $\delta \cV_\alpha \equiv \t\Delta_\alpha \delta V|_{\t\Theta=0}$ and the $\t\Theta^2$ factor  to change the $x^n$ dependence of $J_{\alpha}$ to $\tilde x^n$ since $\t \Delta_{\alpha} \tilde x^n = 0$.%
\footnote{$J_\alpha$ has $x^n$ dependence since the gauge invariant current depends on the 4d $\cV_\alpha$.}
The equations of motion we obtain are
\begin{align} \label{}
- \frac{1}{2} (D^\alpha W_\alpha + \b D_{\dot\alpha} \b W^{\dot\alpha}) = 2\delta(\tilde x^n) \t\Theta^\alpha J_\alpha(x^i, \tilde x^n, \Theta)
\end{align}
with the identification of $\delta(\tilde x^n) \t\Theta^\alpha J_\alpha$ as the embedding of the current in the 4d superspace. We discuss this embedding in more details in the next section. 


\section{Warm-up -- global conserved currents}
\label{sec:current}

In this section we study multiplets of global conserved currents. We reviewed in section \ref{sec:3d} the structure of such multiplets in the case of $\cN=1$ in 3d. They are given by a spinor superfield $J_\alpha$ satisfying $\cD^\alpha J_\alpha=0$. We shall shortly remind the reader of its 4d counterpart. Our goal in this section is to formulate the most general conservation equation which is consistent with the symmetries of a 4d theory interacting with a 3d defect. This is a useful preliminary to our study of current multiplets pertaining to spacetime (superspace) symmetries to which we turn in the next section.

Let us begin by recalling that the 
conserved current multiplet in 4d
is defined as a real multiplet $J$ satisfying $\b D^2 J =0$. To see what 
relations this constraint implies on the components of $J$ let us consider the standard superspace expansion of a real multiplet, given in appendix~\ref{4d_superspace_app}. In terms of these components we find \eqref{chiral_from_V}
\begin{align} \label{4d_current_constraint}
\b D^2 J = 2i \b M + 4 \theta(i\lambda - \sigma^\mu \d_\mu \b \chi) -2 \theta^2 (D+ \d^2 C - i \d_\mu v^\mu),
\end{align}
and imposing the constraint implies $\d_\mu v^\mu =0$ and the following component expansion of $J$, in which we renamed the fields for later convenience,
\begin{align} \label{4d_current}
J = f + i \theta \rho - i \b \theta \b \rho - \theta \sigma^\mu \b \theta j_\mu + \tfrac{1}{2} \theta^2 \b \theta \b \sigma^\mu \d_\mu \rho - \tfrac{1}{2} \b \theta^2 \theta \sigma^\mu \d_\mu \b\rho - \tfrac{1}{4} \theta^2 \b \theta^2 \d^2 f.
\end{align}
Let us note that the constraint $\b D^2 J=0$ sets to zero a \emph{chiral} submultiplet \eqref{4d_current_constraint} of $J$.

Next we show that a 3d current multiplet can be embedded in a 4d real multiplet satisfying the same constraint. As discussed previously, a 3d current resides in a spinor multiplet $J_\alpha$ satisfying $\Delta^\alpha J_\alpha=0$. Since the fields are so far 3d with no $x^n$ dependence we might as well use $\Delta_\alpha$ instead of $\cD_\alpha$. A natural guess for the 4d embedding is 
\begin{align} \label{glob_current_emb}
\t J = \delta(\tilde x^n) \t \Theta^\alpha J_\alpha,
\end{align}
and a simple computation confirms that $\b D^2 \t J = -\frac{i}{2} \delta(\tilde y^n) \b D^2(\t\Theta^2\Delta^\alpha J_\alpha)$. 
This shows that the 3d constraint for $J_\alpha$ is exactly equivalent to the 4d one for $\t J$.
Another way to understand the expression for $\t J$ is to consider the decomposition of $J$ following the prescription given above. We find
\begin{align} \label{}
J|_{\t \Theta =0} &= f + \Theta \kappa - \frac{1}{2} \Theta^2 j^n, \\
\t \Delta_\alpha J|_{\t \Theta =0} &= \chi_\alpha + \Theta_\alpha \d_n f + i(\Gamma^{i} \Theta)_\alpha j_{i} + \frac{i}{2} \Theta^2\left(\Gamma^{i}\d_{i} \chi + 2i \d_n \kappa \right)_\alpha,
\end{align}
where $\kappa_\alpha = \frac{i}{\sqrt{2}} (\rho- \sigma^n \b \rho)_\alpha$ and $\chi_\alpha = \frac{1}{\sqrt{2}} (\rho+\sigma^n \b \rho)_\alpha$. Setting $J|_{\t \Theta =0} =0$ means that $j^{i}$ is conserved in the 3d sense. Moreover, we find that $\t \Delta_\alpha J|_{\t \Theta =0}$ is identical to the expression for the 3d conserved current multiplet \eqref{3d_cons_current}. 

Let us now consider the case where the 3d and 4d theories are coupled. It turns out that in this case the two terms above $J$ and $\t J$ are not sufficient for the constraint to hold. (We illustrate this below in an explicit example.) It is not far-fetched to speculate that what we are missing is a $\t \Theta^2$ term, however there is a more elegant way of discovering this term. 

Looking back at \eqref{4d_current_constraint}, we see that to guarantee a conserved current it is sufficient to constrain the imaginary part of the $\theta^2$ component of the chiral superfield. Normally, a 4d chiral superfield must be complex and hence the constraint above is the minimal possible. However as discussed in the previous section, owing to the coupling with 3d, we have a natural construction of ``real chiral superfields''. We therefore relax the constraint to 
\begin{align} \label{global_current_3d4d}
\b D^2 J = \delta(\tilde y^n) \cB
= \delta(\tilde y^n) \left( b + \sqrt{2} \theta \chi_b + \theta^2(f_b-i\d_n b)\right),
\end{align}
where $b$ and $f_b$ are real and $\chi_b$ is Majorana.
The argument of the delta function is again crucial. Expanding $\tilde y^n = y^n - i \theta^2$, we see that the imaginary part of the $\theta^2$ component is a total derivative. This implies $\d_\mu v^\mu = -\frac{1}{2}\d_n\left( \delta(x^n) b \right)$ and lets us define a conserved current by $j^\mu = v^\mu + \frac{1}{2}\delta^{\mu}{}_n\delta(x^n) b$. The new term in the current is understood in light of the form of the projection in \eqref{43chiral_proj}-\eqref{43a_chiral_proj}. The normal derivatives in $\Phi |_{\t\Theta=0}$ and $\b \Phi |_{\t\Theta=0}$ mean that the potential involves derivative interactions and therefore contributes to the current, as follows from Noether's formula. 

As a final comment, let us show that the new term on the right hand side of \eqref{global_current_3d4d} can be written as a $\t\Theta^2$ contribution to $\t J$, as remarked above. For this purpose, define the projection $B = \cB|_{\t\Theta=0}$. Then using \eqref{chi_emb_formula} we have the equality $\delta(\tilde y^n) \cB = \frac{1}{2} \b D^2 (\delta (\tilde x^n) \t\Theta^2 B)$, demonstrating our claim.

\subsection{A derivation using superspace Noether procedure}
\label{sec:var1}

We now show how the equation for the current can be obtained from a variational approach. Let us start from a global $U(1)$ symmetry. It acts on the matter fields by $\delta \Phi^a = i \omega q_a \Phi^a$ and $\delta A^I = i \omega q_I A^I$, where $\omega$ is the parameter of transformation and $q_{a}$ and $q_I$ are the charges. To obtain the current, the symmetry is gauged by giving a space time dependence to the symmetry parameter. This is implemented in superspace in the following way. In the case of 4d chirals, $\omega$ is lifted to a chiral superfield $\Omega$ by defining $\delta \Phi^a = i \Omega q_a \Phi^a$. The global limit is obtained by equating $\Omega= \b \Omega$. Chiral and anti-chiral fields are equal if and only if all fields vanish except for the real part of the bottom component which has to be constant. This means that the variation of the 4d Lagrangian
takes the form \cite{Osborn:1998qu}
\begin{align} \label{}
\delta \mathscr{L}^{(4)} = \int d^4 \theta i(\Omega- \b \Omega) J
\end{align}
for some $J$. The variation must vanish on the equations of motion for any $\Omega$ and therefore we can obtain $\b D^2 J = D^2 J=0$. Similarly, we introduce in 3d a real multiplet $\omega$ which gauges the symmetry, and the global limit is obtained by $\Delta_\alpha \omega=0$. The Lagrangian hence transforms as
\begin{align} \label{}
\delta \mathscr{L}^{(3)} = \int d^2 \Theta \Delta^\alpha \omega J_\alpha
\end{align}
for some $J_\alpha$. This leads to the conservation equation $\Delta^\alpha J_\alpha=0$.

Let us now assume that the theories are coupled in a supersymmetric way. We make the identification $\omega = \frac{1}{2}(\Omega + \b \Omega)|_{\t\Theta=0}$ and define $\omega' = -\frac{i}{2}(\Omega - \b \Omega)|_{\t\Theta=0}$. Varying the Lagrangian as above we get a new term since $\omega'$ vanishes in the global limit, \emph{i.e.}
\begin{align} \label{var_metho_glob_current}
\delta \mathscr{L}^{\mathrm{tot}} = \int d^4 \theta i(\Omega- \b \Omega) J 
+ \int d^2 \Theta \, \delta(x^n) \left( \Delta^\alpha \omega J_\alpha 
- \omega' B \right).
\end{align}
The 4d part can be written as $-\frac{i}{4}\int d^2 \theta \Omega \b D^2 J + c.c.$ and then projected into the 3d superspace
\begin{align} \label{}
-\frac{i}{2}\int d^2 \Theta (\omega + i\omega') \b D^2 J|_{\t\Theta=0} +\frac{i}{2}\int d^2 \Theta (\omega - i \omega') D^2 J|_{\t\Theta=0}.
\end{align}
We obtain the conservation equations
\begin{align} \label{gen_conserv_eq}
-\frac{i}{2}\left( \b D^2 J - D^2 J\right)|_{\t\Theta=0} = \delta(x^n)\Delta^\alpha J_\alpha, \qquad
\frac{1}{2}\left( \b D^2 J + D^2 J\right)|_{\t\Theta=0} = \delta(x^n) B,
\end{align}
or more conveniently $\b D^2 J |_{\t\Theta=0} = \delta(x^n)(i\Delta^\alpha J_\alpha + B)$. Using the same methods as above this can be lifted to the 4d superspace expression we found in the previous section. 

As an example, consider 4d chirals $\Phi^a$ coupled to 3d scalar multiplets $A^I$ transforming as indicated above. A potential $P(\Phi^a, \b \Phi^{\b a}, A^I, \b A^{\b I})|_{\t\Theta=0}$ is invariant if
\begin{align} \label{}
\delta P = i\sum_a q_a (P_a \Phi^a - P_{\b a} \b \Phi^{\b a}) + i\sum_I q_I (P_I A^I - P_{\b I} \b A^{\b I}) = 0.
\end{align}
After gauging, the fields transform by $\delta \Phi^a|_{\t\Theta=0} = iq_a (\omega+ i \omega') \Phi^a|_{\t\Theta=0}$ and $\delta A^I = i q_I \omega A^I$. This leads to
\begin{align} \label{}
\delta P |_{\t\Theta=0} = - \omega' \sum_a q_a (P_a \Phi^a - P_{\b a} \b \Phi^{\b a})|_{\t\Theta=0} = - \omega' B.
\end{align}
It is a trivial exercise to compute $J$ and $J_\alpha$ assuming some $U(1)$ invariant kinetic terms for $\Phi^a$ and $A^I$ and to show that the conservation equation \eqref{gen_conserv_eq} is satisfied.

\section{Energy-momentum multiplet in 4d}
\label{sec:defect}

The purpose of this section is to suggest a modification of the $\cS$-multiplet that comes from the interaction with a 3d defect preserving $\cN=1$ supersymmetry. We do it in two stages. First, we show how to embed the energy-momentum multiplet of a purely 3d theory, \emph{i.e.} equation \eqref{3d_S_mult}, in the 4d $\cS$-multiplet. This is important since the total energy-momentum tensor should be of the form $T^{(4)}_{\nu\mu}+ \delta(x^n) \cP_\nu{}^i \cP_\mu{}^jT^{(3)}_{ij}$, where $\cP_\mu{}^j$ is the embedding defined in the introduction, with $\cP_n{}^j=0$ and $\cP_k{}^j = \delta_k{}^j$. However, the structure that is obtained is not sufficient to describe the coupling of 4d theory with a 3d theory. We therefore study in section~\ref{sec:subsec} what terms can appear on the right hand side of the $\cS$-multiplet which are consistent with an energy-momentum tensor conserved in the 3 directions tangent to the defect and a conserved Majorana supercurrent. Finally, we elaborate on two examples and compute the resulting displacement operators. 

Let us reiterate here, for convenience, the definitions of the 3 and 4 dimensional energy-momentum multiplets. First, the 4d $\cS$-multiplet \cite{Komargodski:2010rb} is given by 
\begin{align} \label{S_mult2}
\b D^{\dot\alpha} \cS_{\alpha \dot\alpha} = 2(\chi_\alpha - \cY_\alpha),
\end{align}
with $\chi_\alpha$ chiral and $D^\alpha\chi_\alpha = \b D_{\dot\alpha} \b \chi^{\dot\alpha}$ and $\cY_\alpha$ satisfying $\b D^2 \cY_\alpha =0$ and $D_{(\alpha} \cY_{\beta)} =0$. The condition on $\chi_\alpha$ means that it can be solved locally as $- \frac{1}{4} \b D^2 D_\alpha V$ where $V$ is a real multiplet. Similarly, the condition on $\cY_\alpha$ means that it can be solved locally as $D_\alpha X$ with $X$ chiral. In appendix~\ref{app:S} we review the $\cS$-multiplet in more detail, including its component expansion, improvements and some examples. In 3d we found the multiplet \eqref{3d_S_mult}
\begin{align} \label{3d_S_mult2}
\Delta^\alpha \cJ_{\alpha i} = -2\d_{i} \Sigma, \qquad
(\Gamma^{i})_\alpha{}^\beta \cJ_{\beta i}= i \Delta_\alpha (H-\Sigma),
\end{align}
where $\Sigma$ and $H$ are real 3d multiplets. Note that we have replaced $\cD_\alpha$ with $\Delta_\alpha$. As remarked before, on 3d fields their action is identical.

\subsection{Embedding the 3d multiplet}

To determine the way the 3d energy-momentum multiplet sits in the 4d $\cS$-multiplet it is most illuminating to consider its component expansion. Keeping in mind our discussion of global conserved currents (see \eqref{glob_current_emb} and below), it is natural to guess that the 3d multiplet should appear as the $\t \Theta_\alpha$ component of the $\cS$-multiplet. Indeed, this is verified by computing the $\t \Delta_\alpha$ derivative of the component expansion appearing in equation \eqref{S_mult_exp_app}. In a purely 3d theory the normal component (such as $j_n$) and normal derivatives are null 
and we find
\bal \label{}
\t \Delta_\alpha \cS_{j} |_{\t \Theta =0} &= - \frac{1}{\sqrt2}(S_{j} + \sigma^n \b S_{j})_\alpha
+ 2i \big(\Gamma_{j}(\psi + \sigma^n \b\psi) \big)_\alpha 
+ 2\Theta_\alpha \d_{j} (x+\b x) \cr
&\quad - i (\Gamma^{i} \Theta)_\alpha \left( 2T_{ij} - 4\eta_{ij} A + \epsilon_{i j k n} \d^{k} v^n \right)  \\ 
&\quad - \frac{1}{2} \Theta^2 \left( \frac{i}{\sqrt2} 
\big(\Gamma^i \d_i (S_j + \sigma^n \b S_j )\big)_\alpha  
-2 \big(\Gamma_j \Gamma^i \d_i (\psi+ \sigma^n \b \psi )\big)_\alpha \right).
\eal
This expression matches the component expansion of the 3d energy-momentum multiplet \eqref{3d_S_mult_exp}, by identifying $\Sigma=2(X+\b X)|_{\t\Theta=0}$ and $V = -\frac{1}{2} \t\Theta^2 H = -\frac{1}{2} \theta\sigma^n\b\theta \,\eta + \ldots$ (recall $\chi_\alpha = -\frac{1}{4}\b D^2 D_\alpha V$). The latter is implied by identifying $v^n = \frac{1}{2}\eta$ and using the relation $V|_{\theta \sigma^\nu\b\theta} = - v_\nu$. What we have shown is that we can embed $\cJ_{\alpha j}$ as
\bal \label{}
\cS_{\alpha\dot\alpha} &= \t \Theta^\beta \cJ_{\beta\alpha}{}^\gamma \sigma^n_{\gamma\dot\alpha},
\\
\chi_\alpha &= \frac{1}{8} \b D^2 D_\alpha \left( \t \Theta^2 H \right),
\\
\cY_\alpha &= \frac{1}{8} D_\alpha \b D^2 \left(\t \Theta^2 \,\Sigma \right) 
\eal
which solves the 4d $\cS$-multiplet equations \eqref{S_mult2}. 

It is also useful to show this by an explicit computation. One quickly finds%
\footnote{\label{3d_chiral_exp}Actually, that is somewhat of a lie. The computation is quite tedious if one attempts to carry it out by brute force. Therefore, out of consideration for the reader we show how it can be trivialized by a simple trick. The idea is to use \eqref{chiral_emb_exp}, which here gives
$$
\cJ_{\beta i}(x^{i}, \tilde x^n, \Theta) = 
\cJ_{\beta i}(y^{i}, \tilde y^n, \sqrt2 \theta) + i\t\Theta^\alpha \Delta_\alpha \cJ_{\beta i}(x^{i}, \tilde x^n, \Theta) + \cO(\t\Theta^2).
$$
From this we get
$$
\t\Theta^\beta \cJ_{\beta i}(x^{i}, \tilde x^n, \Theta) 
= \frac{1}{2} \t\Theta^\beta \b D^2\left(\t\Theta^2 \cJ_{\beta i}\right)
- \frac{i}{2} \t\Theta^2 \Delta^\beta \cJ_{\beta i}
$$
Here relation \eqref{chi_emb_formula} was used. Applying $\b D^{\dot\alpha}$ now leads to \eqref{4demb_3dmult}.}
\begin{align} \label{4demb_3dmult}
\b D^{\dot\alpha} \left(\t \Theta^\beta \cJ_{\beta\alpha}{}^\gamma \sigma^n_{\gamma\dot\alpha} \right)
= - \frac{i}{2\sqrt2} \b D^2 \left( \t \Theta^2 \cJ_{\beta\alpha}{}^\beta \right) - \frac{i}{2} (\sigma^n \b D)_\gamma \left( \t\Theta^2 \Delta^\beta \cJ_{\beta\alpha}{}^{\gamma} \right).
\end{align}
Notice that the terms in the two parentheses on the right hand side exactly correspond to the two terms in \eqref{3d_S_mult2} and since both multiply $\t \Theta^2$ we can interchange $\cD_\alpha$ with $\Delta_\alpha$ as 
in \eqref{3d_S_mult2}, without restricting the dependence of the operators on $x^n$. 
We now use this relation to obtain
\bal\label{4demb_3dmult_2}
\b D^{\dot\alpha} \left(\t \Theta^\beta \cJ_{\beta\alpha}{}^\gamma \sigma^n_{\gamma\dot\alpha} \right) 
&= \frac{1}{4}\b D^2 D_\alpha \left( \t\Theta^2 (H-\Sigma)\right) + i (\sigma^{i}\b D)_\alpha \d_{i} \left( \t\Theta^2 \Sigma \right)\cr
&= 2 (\chi_\alpha - \cY_\alpha) - i (\sigma^n \b D)_\alpha \d_n \left( \t\Theta^2 \,\Sigma \right). 
\eal
If $\Sigma$ has no $x^n$ dependence, then the last term in the second line drops out and we obtain the result from before. Roughly speaking, this term is responsible for cancelling the $\d_n$ in $\cY_\alpha$. We define 
\begin{align} \label{Yprime_term}
2\cY'_\alpha = \frac{1}{4}D_\alpha \b D^2\left( \t\Theta^2 \, \Sigma \right) + i (\sigma^n \b D)_\alpha \d_n \left( \t\Theta^2 \, \Sigma \right),
\end{align}
which satisfies $2 \b D_{\dot\alpha} \cY'_\alpha = -i \sigma^{i}_{\alpha\dot\alpha} \d_{i} \Sigma$. 
We note the absence of the normal derivative in this expression. Compare this with the corresponding 4d term in \eqref{S_mult2}, which gives $2 \b D_{\dot\alpha} \cY_\alpha = -4i\sigma^{\mu}_{\alpha\dot\alpha} \d_\mu X$.

For the application we want to consider, the 3d contribution should come with a delta function, and hence is defined as
\begin{align} \label{3d_cont_Smult}
\cS^{(3)}_{\alpha\dot\alpha} = \delta(\tilde x^n) \t \Theta^\beta J_{\beta\alpha}{}^\gamma \sigma^n_{\gamma\dot\alpha}.
\end{align}
Note that we can change $\delta(\tilde x^n) \to \delta(\tilde y^n)$ because of the 
$\t\Theta^\beta$ factor. It satisfies
\begin{align} \label{3d_cont_Smult2}
\b D^{\dot\alpha} \cS^{(3)}_{\alpha\dot\alpha} = 2\delta(\tilde y^n)(\chi_\alpha - \cY'_\alpha).
\end{align}
In fact, we can swallow the delta function in $\Sigma$ and $H$, which is most clearly observed in the first line of \eqref{4demb_3dmult_2}. Since $\delta(\tilde y^n)$ is chiral it goes through $\b D_{\dot\alpha}$ with no effect. But we can pull the delta through $D_\alpha$ as well since the commutation relation is proportional to $\t\Theta_\alpha$. 

Lastly, let us note that when considering also the 4d theory, we can form the combination
\begin{align} \label{}
\cS_{\mu} \equiv \cS^{(4)}_{\mu} + \cS^{(3)}_{\mu} = 2\theta\sigma^\nu \b\theta\left(T^{(4)}_{\nu\mu} + \delta(x^n)\cP_\nu{}^i \cP_\mu{}^j T^{(3)}_{ij}\right) + \cdots\,.
\end{align}
Since we showed that the 3d terms can be swallowed in the $\cS$-multiplet terms (the difference between $\cY_\alpha$ and $\cY'_\alpha$ is immaterial), it should be obvious that this can not lead to a displacement operator. In other words, the resulting energy-momentum tensor will be fully conserved, which follows straightforwardly by acting with $\b D_{\dot\alpha}$ on \eqref{3d_cont_Smult2} and noting that the right hand side is a total derivative (see also the discussion around \eqref{quick_disp} below). This means that the structure we have described can not accommodate 4d theories coupled to 3d theories, which requires the appearance of a new term. In the next section we investigate the form such terms can take. 
Note that also in the purely bosonic cases reviewed
in appendix~\ref{app:scalar}, the displacement operator vanishes when the 4d and 3d degrees of freedom are not coupled.

\subsection{The defect multiplet}
\label{sec:subsec}

We would now like to find new terms that can appear on the right hand side of the $\cS$-multiplet and are consistent with the existence of a conserved energy-momentum tensor in the 3 directions parallel to the defect and a conserved Majorana supercurrent (\emph{i.e.} a supercurrent which is a linear combination of 4d supercurrents of opposite chirality). Since the equation for the $\cS$-multiplet is linear and the solution for the 4d terms $\chi_\alpha$ and $\cY_\alpha$ is well known, we might as well discard them and focus on the term of interest to us. We therefore take the following starting point
\begin{equation} \label{def_mult2}
\b D^{\dot\alpha} V_{\alpha\dot\alpha} = 2\delta(\tilde y^n) \cZ_\alpha \equiv 2\cZ'_{\alpha}.
\end{equation}
Here $V_\mu = (C_\mu, \chi_\mu,M_\mu,v_{\nu\mu},\lambda_\mu, D_\mu)$ is a real vector multiplet. It is obtained simply by adding a vector index to the usual real multiplet $V$ discussed in appendix~\ref{app:conv}. Our goal is to find constraints on $\cZ_\alpha$ that lead to a multiplet with the requirements specified above. Consistency requires $\b D^2 \cZ_\alpha = 0$, and we can define a chiral superfield $\Pi_{\alpha\dot\alpha} = -2i \b D_{\dot\alpha} \cZ_\alpha$.

A quick way to show what conditions ensure the existence of a conserved energy-momentum tensor in the directions parallel to the defect and to see how the displacement operator emerges is to follow a similar argument to that which appeared in section~\ref{sec:current}. From \eqref{def_mult2} we can derive $\b D^2 V_\mu = 2i \delta(\tilde y^n)\Pi_\mu$. This is compared to an expression similar to \eqref{4d_current_constraint}
\begin{equation} \label{quick_disp}
\b D^2 V_\mu = \cdots - 2\theta^2 (D_\mu + \d^2 C_\mu - i \d^\nu v_{\nu\mu}).
\end{equation}
We learn from this that the existence of a conserved energy-momentum tensor imposes that the real part of $\Pi_{i}$ is a total derivative while the real part of $\Pi_n$ gives the displacement operator. It should be noted that $v_{\mu\nu}$ is not symmetric here so we ought to be a little careful. In particular the current index in $\cS_\mu$ is the free vector index. 

To argue more systematically, we obtain the following equation
\begin{equation} \label{def_mult3}
\d_\mu V^\mu = -\frac{i}{2} (D^\alpha \cZ'_\alpha - \b D_{\dot\alpha} \b \cZ'^{\dot\alpha}),
\end{equation}
which is derived from \eqref{def_mult2}.
It is useful to look closer at the components of $\d_\mu V^\mu$. In particular the interesting sub-multiplet is given by 
\bal\label{}
\t \Delta_\alpha \d_\mu V^\mu |_{\t\Theta=0} &= \frac{1}{\sqrt2}\d_\mu(\chi^\mu + \sigma^n \b \chi^\mu)_\alpha + \Theta_\alpha \d_\mu \left(\frac{1}{2}(M^\mu+\b M^\mu) + \d_n C^\mu \right) \cr
&\quad + i (\Gamma_j \Theta)_\alpha \d_\mu v^{j\mu}  
- \frac{1}{2} \Theta^2 \d_\mu \left(  \kappa_\alpha^\mu + \frac{i}{\sqrt2}\big(\Gamma^j \d_j(\chi^\mu + \sigma^n \b \chi^\mu)\big)_\alpha\right).
\eal
Here we have introduced $\kappa_{\alpha}^\mu \equiv \sqrt2 (\lambda+ \sigma^n \b \lambda )_{\alpha}^\mu$ only for the sake of keeping the length of the expression in check.
We recognize that this sub-multiplet contains the components that we want to keep conserved, namely $\chi_\mu + \sigma^n \b \chi_\mu$ and $v_{j\mu}$.

Projecting to the same sub-multiplet on the right hand side of \eqref{def_mult3} we find
\beq \label{find_Z_constraints1}
\sqrt2 \t \Delta_\alpha \d_\mu V^\mu |_{\t\Theta=0}= \Delta^\beta \Delta_\alpha (\cZ' +\sigma^n \b \cZ')_\beta
+ \frac{i}{\sqrt2} \Delta_\beta( \Pi'_\alpha{}^\beta + \b \Pi'_\alpha{}^\beta)
+2 i\d_n(\cZ'- \sigma^n \b \cZ')_\alpha.
\eeq
Here $\Pi'_{\alpha}{}^\beta = \delta(\tilde y^n) \Pi_{i} (\Gamma^{i})_\alpha{}^\beta$. Since $\delta(\tilde y^n)|_{\t\Theta=0}= \delta(x^n)$ and the $\d_n$ term in $\Delta_\alpha$ is of order $\t\Theta_\alpha$ we can simplify to 
\bal \label{find_Z_constraints2}
&= \delta(x^n) \left.\left(\Delta^\beta \Delta_\alpha (\cZ +\sigma^n \b \cZ)_\beta
+ \frac{i}{\sqrt2} \Delta_\beta( \Pi_\alpha{}^\beta + \b \Pi_\alpha{}^\beta) \right) \right|_{\t\Theta=0} 
\\&\quad 
+ 2i \d_n \left( \delta(x^n)(\cZ- \sigma^n \b \cZ)_\alpha\right)|_{\t\Theta=0}.
\eal
What are the conditions which guarantee the existence of a conserved energy-momen\-tum tensor and supercurrent? For the first term we can demand that the 3d projection of $\cZ_\alpha + (\sigma^n \b \cZ)_\alpha$ is either a total covariant derivative $\Delta_\alpha (\ldots)$ or it is a current, \emph{i.e.} annihilated by $\Delta^\alpha$. Likewise we demand that the 3d projection of $\Pi_{i}+ \b \Pi_{i}$ is a total derivative $\d_{i}(\ldots)$. The obvious solutions are the ones we already encountered above, namely $\cZ_\alpha = \chi_\alpha$, $\cY_\alpha$ and $\cY'_\alpha$. Let us focus here on a different solution given by imposing
\begin{equation} 
\left(\cZ_\alpha+ (\sigma^n \b \cZ)_\alpha \right) |_{\t\Theta=0} = 0, \qquad 
(\Pi_{\mu} + \b \Pi_{\mu})|_{\t\Theta=0} = -4n_\mu \mathscr{D} \label{demand1}.
\end{equation}
Here $\mathscr{D}$ is a real scalar multiplet (of the 3d superspace), which we now show contains the displacement operator.

To see more explicitly the conservation equation for the energy-momentum tensor we proceed as follows. The $\theta \sigma^\nu\b \theta$ component of $\d_{\mu}V^\mu$ (where the energy-momentum sits) is obtained as the bottom component of
\bal
\label{DDdV}
[D_\alpha, \b D_{\dot\alpha} ] \d_\mu V^\mu 
&=\frac{1}{4} D^2 \left( \delta(\tilde y^n) \Pi_{\alpha\dot\alpha}\right) 
+\frac{1}{4} \b D^2 \left( \delta(\tilde {\b y}^n) \b \Pi_{\alpha\dot\alpha}\right) \cr
&\quad + \d_{\beta \dot\alpha} \left( D_\alpha \cZ'^\beta + D^\beta \cZ'_\alpha \right) 
- \d_{\alpha\dot\beta} \left( \b D^{\dot\beta} \b \cZ'_{\dot\alpha} + \b D_{\dot\alpha} \b \cZ'^{\dot\beta} \right) 
\eal
It is immediate to evaluate the bottom component of the top line by using \eqref{Delta_to_D1}
\begin{align} \label{}
\d^\mu v_{\nu\mu} = -\frac{i}{2} \d_n \left(\delta(x^n) (\Pi_{\nu} - \b \Pi_{\nu}) \right) - \frac{1}{4} \delta(x^n) \Delta^2 (\Pi_{\nu}+ \b \Pi_\nu) + \cdots,
\end{align}
where the ellipses represent the contributions from the second line of \eqref{DDdV}, which are total derivatives, as in fact is also the first term here, so the interesting contribution is the second term
\begin{align} \label{}
\d^\mu v_{n \mu} 
= -2 \delta(x^n) f_d + \cdots,
\end{align}
where $\Delta^2 \mathscr{D}| = -2 f_d$. As we see below, $-2T_{\nu\mu} = v_{\nu\mu}+ \cdots$ which leads us the form of the displacement operator given in \eqref{disp_def}.

\subsection{The components of the defect multiplet}

To find the components of $V_\mu$ we first need to solve the constraints \eqref{demand1} on $\cZ_\alpha$ more explicitly. 
Using a chiral superfield expansion we can write $\cZ_\alpha = i\Lambda_\alpha - \frac{i}{2} (\sigma^\mu \b \theta)_\alpha \Pi_\mu$, where $\Lambda_\alpha$ and $\Pi_\mu$ are chirals (although $\Lambda_\alpha$ does not transform standardly%
\footnote{By this we mean, that the supersymmetry variation of $\Lambda_\alpha$ contains $\Pi_\mu$ terms. This is a consequence of the explicit use of $\b\theta_{\dot\alpha}$ in this definition. The same is not true of $\Pi_\mu$ since it has a natural superspace definition as the covariant derivative of $\cZ_\alpha$.}%
). In fact, it is even more convenient to redefine $\Lambda_\alpha \to \Lambda_\alpha + \frac{1}{2} (\sigma^\mu \b \sigma^n \theta)_\alpha \Pi_\mu$, which leaves $\Lambda_\alpha$ chiral. The advantage is that now
\begin{align} \label{}
\cZ_\alpha = i\Lambda_\alpha - \frac{1}{\sqrt2} (\sigma^\mu \b \sigma^n \t \Theta)_\alpha \Pi_\mu
\end{align}
and $\cZ_\alpha |_{\t\Theta=0} = i\Lambda_\alpha |_{\t\Theta=0}$ (hence $\Lambda_\alpha|_{\t\Theta=0}$ does transform standardly under the preserved subalgebra), while still maintaining the relation $-2i \b D_{\dot\alpha} \cZ_\alpha = \Pi_{\alpha\dot\alpha}$. The components are given by
\bal
\Lambda_\alpha &= \rho_\alpha + \theta_\alpha B - i(\sigma^{\mu\nu}\theta)_\alpha \Lambda_{\mu\nu} + \theta^2 \kappa_\alpha, \cr
\Pi_\mu &= g_\mu + \sqrt{2} \theta \psi_\mu + \theta^2 F_\mu.
\eal
Here $\Lambda_{\mu\nu}$ may be taken to be real. 
We also expand the 3d multiplet $\mathscr{D}$ as
\beq
\mathscr{D} = d + \Theta \chi_d + \frac{1}{2} \Theta^2 f_d.
\eeq
The first constraint in \eqref{demand1} implies 
\beq
\rho = \sigma^n \b \rho, \qquad
\Im(B) = 0, \qquad
\Lambda_{i j} = 0, \qquad 
 \kappa + i\d_n \rho= \sigma^n (\b\kappa - i\d_n \b\rho).
\eeq
We define $\ell_\mu = \Lambda_{n \mu}$.
The second constraint in \eqref{demand1} gives
\bal{}
&\Re(g_{\mu}) = -2n_{\mu} d, \qquad
\psi_{\mu} + \sigma^n \b \psi_{\mu} = -2 n_{\mu} \chi_d, \\
{}&\Re(F_{\mu}) - \d_n \Im(g_{\mu}) = -2n_{\mu} f_d.
\eal

We are now ready to solve \eqref{def_mult2}, with $\cZ_\alpha$ subject to the constraints \eqref{demand1}, by expressing the components of $V_\mu$ ($v_{\mu\nu}$ and $\chi_\mu$) in terms of the conserved quantities $T_{\nu\mu}$ and $S_\mu + \sigma^n \b S_\mu$. As an example, consider taking the bottom component of \eqref{find_Z_constraints1}. Recalling that $V_\mu = C_\mu + i \theta\chi_\mu - i\b\theta \b \chi_\mu + \ldots$ as in \eqref{real_mult_app}, this leads to the relation
\begin{align} \label{}
\d_\mu (\chi^\mu + \sigma^n \b \chi^\mu) 
&= -2 \d_n (\rho' + \sigma^n \b \rho') \cr
&= -2\d_\mu (\sigma^\mu \b \rho' - \sigma^n \b \sigma^\mu \rho').
\end{align}
Here we are again using the shorthand $\rho' = \delta(x^n) \rho$, and in the second line we have used that $\rho$ is a Majorana spinor, namely $\rho = \sigma^n \b\rho$. This allows us to define a conserved supercurrent by
\begin{align} \label{}
\sqrt2 \, \hat S^\mu = - (\chi^\mu + \sigma^n \b \chi^\mu) - 2 \delta(x^n) (\sigma^\mu \b \rho - \sigma^n \b \sigma^\mu \rho).
\end{align}
We can now write $\sqrt2 \, \hat S^\mu = S^\mu + \sigma^n \b S^\mu$ and decompose the relation above to
\begin{align} \label{S_from_chi}
\chi^\mu = - S^\mu -2 \delta(x^n) \sigma^\mu \b \rho, 
\qquad
\b \chi^\mu = - \b S^\mu + 2 \delta(x^n) \b \sigma^\mu \rho, 
\end{align}
noting that $S^\mu$ is here determined only up to a shift by imaginary spinors (\emph{i.e.} $\zeta^\dagger = -\zeta \sigma^n$), such that $\hat S^\mu$ remains unchanged. With a similar analysis of $v_{\mu\nu}$ we get the expansion
\begin{align} \label{}
V_\mu &= C_\mu - i \theta \left( S_\mu + 2 \delta(x^n) \sigma_\mu \b \rho \right) + i \b \theta\left( \b S_\mu - 2\delta(x^n) \b\sigma_\mu \rho \right) + \frac{i}{2}\theta^2 \delta(x^n) \b g_\mu - \frac{i}{2} \b \theta^2 \delta(x^n) g_\mu \nonumber\\
&\quad + \theta \sigma^\nu \b \theta \left( 2 T_{\nu\mu} - \frac{1}{2}\epsilon_{\nu\mu\rho\kappa} \d^\rho C^\kappa -
 \delta(x^n) \left( n_{\nu} \Im(g_\mu) - 4 n_{[\nu} \ell_{\mu]} \right) \right) + \cdots,
\end{align}
where the different fields satisfy the following conservation equations
\begin{align} \label{}
\d^{\mu} T_{\nu\mu} = n_{\nu}\delta(x^n)f_d, \qquad 
\d^\mu (S_\mu +\sigma^n \b S_\mu) = 0.
\end{align}
The violation of conservation of momentum in the normal direction, \emph{i.e.}, the displacement operator, is accompanied by a similar statement for the supercurrent, which takes the form
\begin{align} \label{}
- i\d_\mu (S^\mu - \sigma^n \b S^\mu) 
&= (\kappa' +\sigma^n \b \kappa') + i \d_\mu (\sigma^\mu \b \rho' + \sigma^n \b \sigma^\mu \rho') + 4 \sqrt2 \chi'_d .
\end{align}
This term, which like the displacement operator is localized on the defect, is afflicted by the ambiguity in $S^\mu$ mentioned below \eqref{S_from_chi}, which implies that we can shift this by a total derivative.
In addition, we have the relations
\begin{align} \label{}
T^\mu{}_\mu = 0, \qquad 
\b \sigma^\mu S_\mu = 6\delta(x^n) \b \rho, \qquad
\d^\mu C_\mu = 2\delta(x^n)(d-B),
\end{align}
and lastly the antisymmetric part of the energy-momentum tensor is given by 
\begin{align} \label{}
T_{[n i]} = \frac{1}{4} \delta(x^n) \left(  \Im(g_{i}) - 2\ell_{i} \right), 
\qquad
T_{[ij]} =0.
\end{align}
Let us note that the new term $\cZ_\alpha$ does not contribute to the trace of the energy-momentum tensor. Since the trace of the supercurrent $\b \sigma^\mu S_\mu$ is a Majorana spinor, we can also define a conserved superconformal current by $x_\nu (\sigma^\nu \b S^\mu + \sigma^n \b \sigma^\nu S^\mu)$. Of course, generically the traces receive contributions from $\chi_\alpha$ and $\cY_\alpha$ in \eqref{S_mult2} (as well as the analogous terms coming from 3d) so the conformal currents are not conserved. Here we are only considering the contribution from the new term $\cZ_\alpha$.

\subsection{Example 1 -- scalar multiplets}

In this example there are 4d chiral superfields $\Phi^a$ with K\"{a}hler potential $K$ and superpotential $W$ and 3d real scalar multiplets $A^I$ with target space metric $\cG_{IJ}$. The two theories are coupled through a potential $P(\Phi^a, \b \Phi^{\b a}, A^I)|_{\t\Theta=0}$. As before, we use $\cP$ to denote the chiral embedding of $P$. The equations of motion are
\bal
&\b D^2 K_a = 4 W_a + 2 \delta(\tilde y^n) \cP_a, \cr
&\cD^\alpha(\cG_{IJ} \cD_\alpha A^J) = \frac{1}{2}\d_I \cG_{JK} \cD^\alpha A^J \cD_\alpha A^K + P_I.
\eal
We define the 4d and 3d parts of the energy-momentum multiplet by
\begin{align} \label{}
\cS^{(4)}_{\alpha\dot\alpha} &= K_{a\b a} \b D_{\dot\alpha} \b \Phi^{\b a} D_\alpha \Phi^a, \\
\cS^{(3)}_{\alpha\dot\alpha} 
&= \delta(\tilde x^n) \t \Theta^{\beta} \cJ_{\beta \alpha}{}^\gamma \sigma^n_{\gamma\dot\alpha}
= \delta(\tilde x^n)\t \Theta^{\beta} \left(-2 \cG_{IJ} \d_{i} A^I \Delta_\beta A^J \right) (\Gamma^{i}\sigma^n)_{\alpha\dot\alpha}.
\end{align}
We find for the 4d part
\begin{align} \label{}
&\b D^{\dot\alpha}\cS^{(4)}_{\alpha\dot\alpha} = 2 (\chi_\alpha - \cY_\alpha) - \delta(\tilde y^n) \cP_a D_\alpha \Phi^a, 
\end{align}
where $\chi_\alpha = - \tfrac{1}{4} \b D^2 D_\alpha K$ and $\cY_\alpha = D_\alpha W$. For the 3d part, using identity \eqref{4demb_3dmult}
\begin{align} \label{}
\b D^{\dot\alpha}\cS^{(3)}_{\alpha\dot\alpha} &= 2(\chi_\alpha - \cY'_\alpha) -\frac{1}{2}\delta(\tilde y^n) \cP_I D_\alpha \cA^I.
\end{align}
We can write the new terms as $2(\cZ_\alpha - \delta \cY'_\alpha)$ with
\bal
2\cZ_{\alpha} &= -\frac{1}{2}\left(\cP_a D_\alpha \Phi^a - \cP_{\b a} D_\alpha \t{\b \Phi}{}^{\b a} \right) - \sqrt{2} \t\Theta_\alpha \d_n \cP, \cr
2\delta \cY'_\alpha &= \frac{1}{4}D_\alpha \cP + i (\sigma^n \b D)_\alpha (\t\Theta^2 \cP).
\eal
The second term has the form of $\cY'_\alpha$ in \eqref{Yprime_term} and can be absorbed in it. $\cZ_\alpha$ satisfies
\begin{align} \label{}
\left(\cZ_\alpha + (\sigma^n \b \cZ)_\alpha \right)|_{\t \Theta =0} 
= 0, \qquad
\left(\Pi_\mu + \b \Pi_\mu \right)|_{\t \Theta =0} = -2n_{\mu} \d_n P.
\end{align}
In particular $2\mathscr{D} = \d_n P$. This is the obvious supersymmetric generalization of the scalar expressions in \eqref{scalar-dis}.

\subsection{Example 2 -- gauge interactions}

In this model we have a 4d Abelian gauge field $W_\alpha$ coupled to a 3d matter field $A$. The equations of motion are
\begin{align} \label{}
&D^\alpha W_\alpha = \b D_{\dot\alpha} \b W^{\dot\alpha} = 2 \delta(\tilde x^n)\t\Theta^\alpha J_\alpha, \\
& \mathscr{D}^2 A = \mathscr{D}^2 \b A = 0.
\end{align}
The two parts of $\cS_\mu$ are
\begin{align} \label{}
\cS^{(4)}_{\alpha\dot\alpha} 
&= -2 \b W_{\dot\alpha} W_\alpha, 
\\
\cS^{(3)}_{\alpha\dot\alpha} 
&= \delta(\tilde x^n) \t\Theta^{\beta}\left(-2 \mathscr{D}_{i}A\mathscr{D}_\beta \b A -2 \mathscr{D}_{i}\b A\mathscr{D}_\beta A \right)(\Gamma^{i}\sigma^n)_{\alpha\dot\alpha},
\end{align}
satisfying
\bal
\b D^{\dot\alpha} \cS^{(4)}_{\alpha\dot\alpha} &= 4 \delta(\tilde y^n)\t\Theta^\beta J_\beta W_\alpha, \cr
\b D^{\dot\alpha} \cS^{(3)}_{\alpha\dot\alpha} &= 2(\chi_\alpha - \cY'_\alpha) + \sqrt{2}i\delta(\tilde y^n) (\Gamma^{i}\t\Theta)_\alpha J \Gamma_{i} \cW.
\eal
It should be noted that $J_\alpha$ and $\cW_\alpha$ both represent the chiral embedding of the corresponding 3d fields.%
\footnote{For the first equation it follows from the fact that $J_\alpha$ is a current (satisfying $\Delta^\alpha J_\alpha =0$). Then reasoning similar to those in footnote~\ref{3d_chiral_exp} show that in $\t\Theta^\alpha J_\alpha$ we can take $J_\alpha$ to be the chiral embedding.}
We identify $\cZ^{(4)}_{\alpha} = 2\t\Theta^\beta J_\beta W_\alpha$ and $\cZ^{(3)}_{\alpha} = \frac{i}{\sqrt{2}} (\Gamma^{i}\t\Theta)_\alpha J \Gamma_{i} \cW$. Obviously $\cZ_\alpha |_{\t\Theta=0} =0$ so our first condition for $\cZ_\alpha$ is trivially satisfied. We can then find
\bal
\Pi^{(4)}_\mu + \b \Pi^{(4)}_\mu &= -2 n_{\mu} J \t \cW + \cP_\mu{}^i \, 2i J \Gamma_{i} \cW, \cr
\Pi^{(3)}_{\mu} + \b \Pi^{(3)}_{\mu} &= - \cP_\mu{}^i \, 2i J \Gamma_{i}\cW.
\eal
The displacement multiplet is therefore $2\mathscr{D} = J \t\cW$ and 
({\it c.f.} \eqref{gague-dis})
\begin{align} \label{}
f_d = j^{i} F_{i n}+ \text{fermions}.
\end{align}

\section{Superspace Noether approach to energy-momentum multiplets}
\label{sec:var}

\subsection{4d multiplets}

We now implement the Noether procedure as an alternative method of deriving the energy-momentum multiplet. This was considered by several authors. Our discussion here is mostly based on \cite{Osborn:1998qu,Magro:2001aj} (see also \cite{Buchbinder:1998qv,Kuzenko:2010ni}). To do this, we must promote supersymmetry to a local symmetry, so we consider the set of chirality preserving diffeomorphisms of superspace
\bal
&\delta y^\mu = v^\mu(y, \theta), 
\quad
&& \delta \b y^\mu = \b v^\mu(\b y, \b\theta),  \\ 
& \delta \theta^\alpha = \lambda^\alpha(y, \theta), 
&& \delta \b\theta^{\dot\alpha} = \b\lambda^{\dot\alpha}(\b y, \b\theta), 
\eal
On chiral functions of superspace this corresponds to the differential operator
\bal
&\CL_+ = v^\mu \d_\mu +\lambda^\alpha \frac{\d}{\d \theta^\alpha} 
= h^\mu \d_\mu + \lambda^\alpha D_\alpha, \\
& h^\mu \equiv v^\mu(y, \theta) +2i \b \theta \b\sigma^\mu \lambda(y, \theta).
\eal
Similarly, for anti-chiral functions $\CL_- = \b h^\mu \d_\mu + \b\lambda_{\dot\alpha} \b D^{\dot\alpha}$.
By definition, the action $\CL_+$ preserves chirality $[\b D_{\dot \alpha}, \CL_+]=0$ and $[D_\alpha, \CL_-]=0$. We have the relation
\begin{align} \label{}
\b D_{\dot \alpha} h^\mu = -2i (\lambda \sigma^\mu)_{\dot \alpha}, \qquad
D_\alpha \b h^{\mu} = 2i (\sigma^\mu \b \lambda)_\alpha. 
\end{align}
Hence $\lambda$ and $\b \lambda$ are determined by $h^\mu$ and $\b h^\mu$, which are free except for the constraint
\begin{align} \label{}
\b D^{(\dot \beta} h^{\dot \alpha) \alpha} = 0, \qquad
D_{(\beta} \b h_{\alpha) \dot \alpha} = 0.
\end{align}
This in particular means that we can write $h^{\dot\alpha\alpha} = -2 i \b D^{\dot \alpha} L^\alpha$ and $\b h^{\dot \alpha \alpha} = -2iD^\alpha \b L^{\dot \alpha}$ for an unconstrained superfield $L_\alpha$
\begin{align} \label{gauge_par}
L_\alpha = \ell_\alpha - \frac{i}{2}(\sigma^\mu \b \theta)_\alpha v_\mu + \b \theta^2 \lambda_\alpha,
\end{align}
where $\ell_\alpha$ is an irrelevant chiral superfield, since the gauge transformation is given in terms of $\b D_{\dot \alpha} L_{\alpha}$. 

Recall how in the case of global symmetries in section~\ref{sec:var1}, the gauging involves promoting the global (Abelian) transformation $\delta \Phi = i \Phi$ to an action by a chiral superfield $\Omega$ given by $\delta \Phi = i \Omega \Phi$. The global limit is then obtained by taking $\Omega = \b \Omega$, and implies that the current $J$ should appear in the variation of the Lagrangian as $i(\Omega - \b \Omega)J$. In the same spirit, the basic assertion is that the global limit is given by
\begin{align} \label{superconf_eq}
h^\mu = \b h^\mu.
\end{align}
More precisely, this equation is equivalent to the superconformal Killing equations \cite{West:1997vm}. For example, letting $v_{\mu}| = \epsilon_{\mu}+ i b_{\mu}$ and $-\frac{1}{2}D^\alpha \lambda_{\alpha}| = \Lambda_{(1)} + i\Lambda_{(2)}$, one can verify that \eqref{superconf_eq} implies
\begin{align} \label{conf_killing_eq}
\d_\mu \epsilon_{\nu} + \d_\nu \epsilon_{\mu} = 4\eta_{\mu\nu} \Lambda_{(1)}.
\end{align}
In other words, \eqref{superconf_eq} imposes the conformal Killing vector equation on $\epsilon_\mu$. 
For a superconformal theory the variation of the Lagrangian must assume the form
\begin{align} \label{superconf_mult}
\delta \mathscr{L}^{(4)} = -\frac{i}{2} \int d^4 \theta (h^\mu - \b h^\mu) \cS_\mu = \frac{1}{2}\int d^4 \theta (\b D^{\dot\alpha} L^\alpha - D^\alpha \b L^{\dot\alpha}) \cS_{\alpha\dot\alpha}.
\end{align}
Indeed, expanding in components (and remembering that $\cS_\mu = 2\theta \sigma^\nu\b\theta T_{\nu\mu} + \ldots$) one finds terms such as $\delta \mathscr{L}^{(4)} = - \d^\nu \epsilon^\mu T_{\nu\mu}$.
Since $L_\alpha$ is not constrained, we obtain the superconformal multiplet $\b D^{\dot\alpha} \cS_{\alpha\dot\alpha} = 0$. 
 
To obtain the other multiplets, we need to constrain the gauge parameters $L_\alpha$. The $\cR$-constraint is given by further imposing $D^\alpha \b D^2 L_\alpha+ \b D_{\dot\alpha} D^2 \b L^{\dot\alpha} =0$, which by using \eqref{gauge_par} gives $D^\alpha\lambda_\alpha + \b D_{\dot\alpha} \b \lambda^{\dot\alpha}=0$. This implies $\Lambda_{(1)} = 0$ so \eqref{conf_killing_eq} now reduces to the Killing vector equation of flat space. The FZ-constraint is given by imposing that the chiral superfield $\sigma \equiv - \frac{1}{4} \b D^2 D^\alpha L_\alpha$ vanishes. In terms of fields in \eqref{gauge_par} this reads $\sigma = \d^\mu h_\mu + D^\alpha \lambda_\alpha$. The bottom component of $\sigma$ gives $\d_\mu \epsilon^\mu = 2\Lambda_{(1)}$, which together with \eqref{conf_killing_eq} once more means that $\Lambda_{(1)}=0$. Finally, the $\cS$-constraint is obtained by imposing both conditions.%
\footnote{It follows from \cite{Komargodski:2010rb} that supersymmetric field theories are generally consistent only with the $\cS$-constraint, \emph{i.e.} the smallest gauge symmetry. Additional assumptions are needed to consider the more general gauge symmetries. For example, only superconformal theories can accommodate the gauge symmetry with $L_\alpha$ unconstrained, otherwise we are gauging a broken symmetry.}

We can use two strategies for applying the constraints to the Noether procedure. The first, as presented above, is to think of the constraints as applying to the gauge symmetry itself. Then, varying the action we again obtain \eqref{superconf_mult} but this time since $L_\alpha$ is constrained not all the component of $\b D^{\dot\alpha} \cS_{\alpha\dot\alpha}$ vanish. A slightly more convenient approach is to take $L_\alpha$ unconstrained and think of the $\cR$- and FZ-constraints as part of the global limit, so on the same footing as \eqref{superconf_eq}. Using this point of view for the Noether procedure, we expect additional terms to appear in the variation of the Lagrangian. The most general variation with the $\cR$-constraint is
\beq \label{R_mult_var}
\delta \mathscr{L}^{(4)} = -\frac{i}{2}\int d^4 \theta(h^\mu - \b h^\mu) \cS_\mu 
- \frac{1}{4} \int d^4 \theta \left( D^\alpha \b D^2 L_\alpha + \b D_{\dot\alpha} D^2 \b L^{\dot\alpha} \right) V, \eeq
where $V$ is a real multiplet. 
Since $L_\alpha$ is not constrained, this leads to the $\cR$-multiplet $\b D^{\dot\alpha} \cS_{\alpha\dot\alpha} = 2\chi_\alpha$ with $\chi_\alpha = -\frac{1}{4} \b D^2 D_\alpha V$ a chiral satisfying $D^\alpha \chi_\alpha = \b D_{\dot\alpha} \b \chi^{\dot\alpha}$. For the gauged FZ-symmetry the most general variation of the Lagrangian is 
\beq \label{FZ_mult_var}
\delta \mathscr{L}^{(4)}= -\frac{i}{2}\int d^4 \theta(h^\mu - \b h^\mu) \cS_\mu - \int d^2 \theta\sigma X - \int d^2 \b\theta \b \sigma \b X, 
\eeq
where $X$ is a chiral superfield. This leads to the FZ-multiplet $\b D^{\dot\alpha} \cS_{\alpha\dot\alpha} = -2\cY_\alpha$ with $\cY_\alpha = D_\alpha X$. For the $\cS$-multiplet we simply impose both constraints to get $\b D^{\dot\alpha} \cS_{\alpha\dot\alpha} =2(\chi_\alpha -\cY_\alpha)$.

Let us consider the example of chiral superfields $\Phi^a$ with a K\"{a}hler potential $K(\Phi^a, \b \Phi^{\b a})$ and superpotential $W(\Phi^a)$. The action of the gauge symmetry on a chiral superfield is given by \cite{Osborn:1998qu} 
\beq \label{fz_trans}
\delta \Phi^a = \mathcal{L}_+ \Phi^a.
\eeq
For the FZ-constraint we find up to total derivatives%
\footnote{The following identities, derived from the definitions of $h^\mu$ and $\lambda_\alpha$, are useful
$$
\d_\mu h^\mu = - \frac{i}{12}[D_\alpha, \b D_{\dot\alpha}] h^{\dot\alpha\alpha} + \frac{4}{3}\sigma, \qquad
\d_\mu \b h^\mu = \frac{i}{12}[D_\alpha, \b D_{\dot\alpha}] \b h^{\dot\alpha\alpha} + \frac{4}{3}\b\sigma.
$$
}
\bal
\delta K
&= \frac{i}{4} (h - \b h)^{\dot\alpha\alpha} \left(K_{a\b a} \b D_{\dot\alpha} \b \Phi^{\b a} D_\alpha \Phi^a + \frac{1}{3} [D_\alpha, \b D_{\dot\alpha}] K \right) - \frac{1}{3}(\sigma + \b \sigma) K \\
\delta W &= - \sigma W.
\eal
Comparing this with \eqref{FZ_mult_var} leads to the FZ-multiplet, which is reviewed in \eqref{FZ_for_kahler}.

If the theory has an $R$-symmetry we can also consider the $R$-constraint. Let $R_a$ be the $R$-charges of $\Phi^a$. $R$-invariance implies the relations
\begin{align} \label{}
\sum i(R_a K_a \Phi^a - R_a K_{\b a} \b\Phi^{\b a}) =0, \qquad
\sum R_a W_a \Phi^a = 2W.
\end{align}
It follows from the first relation that $U_\mathcal{R} = \frac{1}{2}\sum R_a K_a \Phi^a$ is a real multiplet. The gauge transformation of a chiral superfield is now \cite{Osborn:1998qu,Magro:2001aj,West:1997vm}
\begin{align} \label{}
\delta \Phi^a = \mathcal{L}_+ \Phi^a + \frac{1}{2} \sigma R_a \Phi^a.
\end{align}
Here there is no sum over $a$. It is easy to check that $\delta W = 0$ up to a total derivative under this gauge symmetry. For the K\"{a}hler potential we find
\bal
\delta K &= \frac{i}{4}(h^{\dot \alpha \alpha} - \b h^{\dot \alpha \alpha}) \left( K_{n\b n} \b D_{\dot \alpha} \b \Phi^{\b n}D_\alpha \Phi^n + [D_\alpha, \b D_{\dot \alpha}] U_\cR \right) \cr
&\quad + (D^\alpha \lambda_\alpha + \b D_{\dot\alpha} \b \lambda^{\dot\alpha})(K - 3 U_\cR). 
\eal
Comparing with \eqref{R_mult_var} one can derive the $R$-multiplet which agrees with \eqref{R_for_kahler}. Finally, for the $\cS$-constraint with \eqref{fz_trans} we find
\bal \label{}
\delta K 
&= \frac{i}{4}(h^{\dot \alpha \alpha} - \b h^{\dot \alpha \alpha}) \left( K_{n\b n} \b D_{\dot \alpha} \b \Phi^{\b n}D_\alpha \Phi^n \right) 
+ ( D^\alpha \lambda_\alpha + \b D_{\dot\alpha} \b \lambda^{\dot\alpha} ) K, \cr
\delta W 
&= - \sigma W,
\eal
which gives the $\cS$-multiplet in \eqref{S_for_kahler}.

\subsection{3d multiplets}

In 3d we consider diffeomorphisms of superspace (see \cite{Park:1999cw} for a related discussion in the superconformal case)
\begin{align} \label{}
\delta x^{i} = \check v^{i}, \qquad 
\delta \Theta^\alpha = K^\alpha.
\end{align}
The action on (scalar) superfields is given by 
\bal{}
\hat{\mathcal{L}}& =\check v^{i} \d_{i} + K^\alpha \frac{\d}{\d \Theta^\alpha} = K^{i} \d_{i} + K^\alpha \cD_\alpha, \\
K^{i} &= \check v^{i} + i \Theta \Gamma^{i} K.
\eal
Contrary to the 4d case, this does not imply any relation between $K^{i}$ and $K^\alpha$, in other words the two superfields are unconstrained. It is not difficult to check that the equation
\beq\label{3d_superconf_eq}
\cD_{\alpha} K^{i} = 2i (\Gamma^{i} K)_\alpha
\eeq
corresponds to the superconformal Killing equation in 3d. To obtain the super Poincar\'e Killing equation we can constrain the gauge symmetry by $\d^{i} K_{i} + \cD^\alpha K_\alpha =0$. Together with \eqref{3d_superconf_eq} this implies also $\cD^{\alpha} K_\alpha=0$.

A general variation of the Lagrangian therefore takes the form
\beq\label{3d_diff_var_action}
\delta \mathscr{L}^{(3)} = -\frac{1}{2} \int d^2 \Theta( \cD^\alpha K^{i} + 2i (K\Gamma^{i})^\alpha) \cJ_{\alpha i}
+ \int d^2 \Theta( - (\d^{i} K_{i} + \cD^\alpha K_\alpha )\Sigma + \cD^\alpha K_\alpha H).
\eeq
Since $K^{i}$ and $K^\alpha$ are unconstrained we readily get the 3d multiplet \eqref{3d_S_mult}.
As an example we consider a sigma model of real scalar multiplets $A^I$ with kinetic term $V_M = \frac{1}{2} \cG_{IJ} \cD^\alpha A^I \cD_\alpha A^J$ and potential $P(A^I)$. The action of the gauge symmetry on $A^I$ is
\begin{align} \label{}
\delta A^I = \hat{\mathcal{L}} A^I = K^{i} \d_{i} A^I + K^\alpha \cD_\alpha A^I.
\end{align}
A simple computation gives
\bal
\delta V_M &= - \frac{1}{2} \left( \cD^\alpha K^{i} + 2i (K \Gamma^{i})^\alpha \right) ( -2 \cG_{IJ} \d_{i} A^I \cD_\alpha A^J)
\\&\quad 
- (\d_{i} K^{i} + \cD^\alpha K_\alpha ) V_M - \cD^\alpha K_\alpha V_M, \cr
\delta P &= - (\d_{i} K^{i} + \cD^\alpha K_\alpha ) P.
\eal
Comparing with \eqref{3d_diff_var_action} leads to the desired form \eqref{sigma_model_mult}.

Let us identify the 3d parameters with the 4d ones. We have
\bal
\check v^{i} &= \frac{1}{2}(\delta y^{i} + \delta \b y^{i})|_{\t\Theta=0} = \frac{1}{2}(v^{i} + \b v^{i})|_{\t\Theta=0}, \cr 
K &= \frac{1}{\sqrt2}(\delta \theta + \sigma^n \delta \b\theta)|_{\t\Theta=0} = \frac{1}{\sqrt2}(\lambda+ \sigma^n \b\lambda)|_{\t\Theta=0}.
\eal
Additionally, 
\begin{align} \label{}
&-\frac{i}{2}(h^{i} - \b h^{i}) = \t K^{i} - \t \Theta^\alpha \left( \Delta_\alpha K^{i} - 2i (\Gamma^{i}K)_\alpha \right) + \cO(\t\Theta^2), \label{diff_34relation} \\
&\t K^{i} = - \frac{i}{2}(v^{i} - \b v^{i}) + i \Theta \Gamma^{i} \t K, \qquad
\t K_\alpha = - \frac{i}{\sqrt2}(\lambda - \sigma^n \b \lambda)_\alpha. 
\end{align}
Clearly $\delta \t \Theta_\alpha = - \t K_\alpha$. We can see that the 4d global limit $h^{i} = \b h^{i}$ corresponds to the 3d equation \eqref{3d_superconf_eq} but includes an additional condition $\t K^{i} =0$. Similarly, for the normal component we find
\begin{align} \label{}
&-\frac{i}{2}(h^n - \b h^n) = \t K^n - \t \Theta^\alpha \left( \Delta_\alpha K^n - 2i \t K_\alpha \right) + \cO(\t\Theta^2), \label{diff_34relation2} \\
&\t K^n = - \frac{i}{2}(v^n - \b v^n) - \Theta K, \qquad
K^n = \frac{1}{2}(v^n + \b v^n) + \Theta \t K. 
\end{align}
The constraints on the gauge symmetry also match 
\begin{align} \label{}
\frac{1}{2}\left(\d^\mu h_\mu + \d^\mu \b h_\mu+ D^\alpha \lambda_\alpha + \b D_{\dot\alpha} \b \lambda^{\dot\alpha} \right)|_{\t\Theta=0} = \d^{\mu} K_{\mu} + \Delta^\alpha K_\alpha.
\end{align}

We now use \eqref{diff_34relation} to rewrite the first term of \eqref{3d_diff_var_action} in the 4d superspace as
\begin{align} \label{}
-\int d^4 \theta \,\t \Theta^\beta \left( \cD_\beta K^{i} - 2i (\Gamma^{i}K)_\beta \right)\delta(\tilde x^n) \t\Theta^\alpha \cJ_{\alpha i} 
&= - \frac{i}{2} \int d^4 \theta (h^{i} - \b h^{i} )\delta(\tilde x^n) \t\Theta^\alpha \cJ_{\alpha i}.
\end{align}
Here we have discarded the term in $h- \b h$ which is of zeroth order in $\t\Theta$ (see \eqref{diff_34relation}) since it does not contribute to the integral. Evidently, this represents a contribution to $\cS_\mu$ in the form $\delta(\tilde x^n)\t\Theta^\alpha \cJ_{\alpha i}$ in agreement with \eqref{3d_cont_Smult}. Similarly, the terms in the second line of \eqref{3d_diff_var_action} are rewritten as
\begin{align} \label{}
&-\int d^2 \Theta (\d^{i} K_{i} + \cD^\alpha K_\alpha )\Sigma
= \int d^4 \theta L^\alpha \left( \frac{1}{8} D_\alpha \b D^2 + \frac{i}{2} (\sigma^n \b D)_\alpha \d_n\right) \left( \delta(\tilde x^n) \t \Theta^2 \Sigma \right)  + c.c.,\nonumber\\
&\int d^2 \Theta\cD^\alpha K_\alpha H 
= \frac{1}{8}\int d^4 \theta L^\alpha \b D^2 D_\alpha\left( \delta(\tilde x^n) \t\Theta^2 H \right) + c.c.,
\end{align}
where in the first line we used $\d^{i} K_{i} + \cD^\alpha K_\alpha= \frac{1}{2}(\sigma + \b \sigma) - \d_n K^n$. This clearly confirms the structure we have found for embedding the 3d energy-momentum multiplet in the $\cS$-multiplet. 

\subsection{The defect multiplet}

Finally, let us see how to obtain the defect multiplet from a variation approach. We proceed by arguments similar to those appearing in the discussion of global currents, see \eqref{var_metho_glob_current}. It follows from \eqref{diff_34relation}-\eqref{diff_34relation2} that in the global limit $\t K^\mu$ and $\Delta_\alpha K^n - 2i \t K_\alpha$ vanish. Moreover, we must demand the vanishing of $K^n$ as well. This guarantees that the solutions to the Killing equations will not include the translation corresponding to the normal direction (and associated transformations). This discussion leads to the following additional terms in the variation of the Lagrangian
\begin{align} \label{var_disp}
\int d^2 \Theta \left( - i \t K^{\mu} \Pi_\mu + 2\sqrt2  \t K^\alpha \Lambda_\alpha + 2 K^n \mathscr{D} \right) . 
\end{align}
Clearly $\Pi_\mu$ must be imaginary while $\Lambda_\alpha$ and $\mathscr{D}$ are real. The dependence on the tilde fields and $K^n$ implies that such terms come from interactions of 4d fields localized on the defect. There terms can be rewritten as
\begin{align} \label{}
\int d^4 \theta L^\alpha \delta(\tilde y^n) \left( i\Lambda_\alpha - \frac{1}{\sqrt2} (\sigma^\mu \b \sigma^n \t \Theta)_\alpha \Pi_\mu \right) + c.c.~,
\end{align}
where we have redefined $(\Pi_n - 2\mathscr{D}) \to \Pi_n$. After this redefinition $\Pi_n$ has a real part which gives the displacement multiplet. Note that, as one can verify by following the derivation, in this equation $\Lambda_\alpha$ and $\Pi_\mu$ are the chiral embeddings of the fields introduced in \eqref{var_disp}. 

As an example consider 4d chiral superfields $\Phi^a$ coupled to real 3d scalar multiplet $A^I$ through a potential $P(\Phi^a,\b \Phi^{\b a}, A^I)|_{\t\Theta=0}$. Projecting the transformations of the 4d chiral to 3d gives
\begin{align} \label{}
\delta \Phi = (K^\mu + i \t K^\mu )\d_\mu \Phi + (K^\alpha + i \t K^\alpha) \Delta_\alpha \Phi. 
\end{align}
Applying this to the interaction potential leads to 
\bal
\delta P &= -( \d_{i} K^{i} + \Delta^\alpha K_\alpha ) P + K^n \d_n P + i \t K^\mu (P_a \d_\mu \Phi^a - P_{\b a} \d_\mu \b \Phi^{\b a} )\cr
&\quad + i \t K^\alpha (P_a \Delta_\alpha \Phi^a - P_{\b a} \Delta_\alpha \b \Phi^{\b a} )
\eal
up to a total derivative. The first term clearly gives rise to a $\cY'_\alpha$ term. We can further obtain 
\bal
\cZ_\alpha &= - \frac{1}{4\sqrt2} \b D^2 \left( \t\Theta^2(P_a \Delta_\alpha \Phi^a - P_{\b a} \Delta_\alpha \t{\b \Phi}{}^{\b a}) \right)
\\&\quad 
- \frac{1}{\sqrt2} (\sigma^\mu \b \sigma^n \t\Theta)_\alpha \left(P_a \d_\mu \Phi^a - P_{\b a} \d_\mu \t{\b \Phi}{}^{\b a} + n_{\mu} \d_n P\right).
\eal
Recall that $\t{\b \Phi}$ is the chiral lift of $\b \Phi|_{\t\Theta=0}$.
This gives $\mathscr{D} = \frac{1}{2} \d_n P$ and matches the results obtained previously. 

\section{Concluding remarks}
\label{sec:outlook}

In this note we discussed 4d ${\cal N}=1$ supersymmetric field theories in the presence of a 3d planar defect, preserving half of the supersymmetry. In particular, we described  how  the displacement operator in these theories is contained in a modified  energy-momentum multiplet, which we named the defect multiplet.  Our main motivation for this work is to understand systematically how to place defects on curved manifolds in a supersymmetric fashion. It will be interesting to develop a formalism that addresses this issue using ideas similar to \cite{Festuccia:2011ws}. A related problem concerns the study of manifolds with boundaries, where one would like to find all possible supersymmetric boundary geometries arising from the rigid limit of background supergravity studied on a manifold with boundaries \cite{Belyaev:2005rt,Belyaev:2007bg,Belyaev:2008ex,Andrianopoli:2014aqa,DiPietro:2015zia,Aprile:2016gvn}.
 
It would be nice to understand the moduli space of all supersymmetric defects and the geometry that characterizes such embeddings. This can then be applied to localization computations and can shed light on the problem of mapping defects and boundaries under dualities.\footnote{Some example of exact results in supersymmetric field theories in the presence of defects include \cite{Lamy-Poirier:2014sea,Bullimore:2014upa,Gomis:2016ljm,Pan:2016fbl}.} In particular, it would be interesting to follow the dependence of the partition function on the moduli space (as in \cite{Closset:2013vra,Closset:2014uda}). It is possible that these methods may also help the study of configurations defined on manifolds with (conformal) boundaries \cite{David:2016onq,Assel:2016pgi}, or be useful for developing a supersymmetric formulation of holographic renormalization \cite{Genolini:2016ecx}.

It will also be interesting to generalize our results to other defects in various dimensions and extended supersymmetry. These include co-dimension two defects in 4d ${\cal N}=1$ field theories, preserving $(0,2)$ supersymmetry in two dimensions, as well as starting from $\cN=2$ in 4d (see \cite{Gaiotto:2013sma} for early work in this direction). The representation of the displacement multiplet for 3d defects preserving $\cN=4$ supersymmetry was studied in \cite{carlo}.

\section*{Acknowledgments}

We are grateful to Benjamin Assel, Stefano Cremonesi, Cristian Vergu and Daisuke Yokoyama for useful discussions, and especially to Cyril Closset, Lorenzo Di Pietro, and Zohar Komargodski  for comments on the manuscript. 
The work of D.M. and I.S. is  supported by the ERC Starting Grant N. 304806,   ``The gauge/gravity duality and geometry in string theory''. The work of N.D. is supported by Science \& Technology Facilities Council via the consolidated grant number ST/J002798/1.


\appendix 

\section{The displacement operator in scalar and gauge field theory}
\label{app:scalar}

Consider a 4d scalar $\phi$ and a 3d scalar $a$, confined to a planar submanifold $\Sigma$. The 
4d and 3d actions are
\bal
\int \mathscr{L}^{(4)}
&=\int \left(-\frac12\partial^\mu\phi\partial_\mu\phi-V_4(\phi)\right),
\\
\int_\Sigma \mathscr{L}^{(3)}
&=\int_\Sigma\left(-\frac12\partial^ia\partial_ia-V_3(a)\right).
\eal
To make the system interesting, we need to couple the 3d and 4d fields. The simplest 
way to do that is
\beq
\int_\Sigma \mathscr{L} ^{(I)}
=-\int_\Sigma V_I(\phi,a),
\eeq
with an arbitrary coupling potential $V_I$.

There are 4d and 3d terms in the energy-moment tensor
\bal
T^{(4)}_{\mu\nu}
&=\partial_\mu\phi\partial_\nu\phi + \eta_{\mu\nu} \mathscr{L}^{(4)}
\\
T^{(3)}_{ij}
&=\partial_ia\partial_ja + \eta_{ij}(\mathscr{L}^{(3)}+\mathscr{L}^{(I)}).
\eal
The full energy-momentum tensor will include both parts, which requires the embedding 
$\cP_\mu{}^i$ on the directions tangent to $\Sigma$
\beq \label{Tmn_comb_app}
T_{\mu\nu}=T^{(4)}_{\mu\nu}+\delta(x^n)\cP_\mu{}^i\cP_\nu{}^jT^{(3)}_{ji}.
\eeq
Using the classical equations of motion we find
\bal
\partial^\mu T_{\mu\nu} = n_\nu \delta(x^n)\partial_\phi V_I(\phi,a)\,\partial_n\phi,
\eal
where $x^n$ is the coordinate normal to $\Sigma$. 
The displacement operator, defined in \eqref{disp_def}, is therefore given by
\beq
\label{scalar-dis}
f_d=\partial_n V_I(\phi,a).
\eeq

In the presence of a 4d Abelian gauge field, the 4d action contains the term
\beq
\int \mathscr{L}^{(4)}=-\frac{1}{4}\int F_{\mu\nu}F^{\mu\nu}.
\eeq
This can couple to a 3d theory on the defect, by gauging a global $U(1)$ symmetry, with current $j_{(3)}^k$, via the coupling
\beq
\label{gauge-coupling}
\int \mathscr{L}^{(I)}=\int_\Sigma v_\mu \cP^\mu{}_k j_{(3)}^k.
\eeq
The bulk energy-momentum is
\beq
T^{(4)}_{\mu\nu}
=F_{\mu\rho}F^{\rho}{}_\nu+ \eta_{\mu\nu}\mathscr{L}^{(4)},
\eeq
and the 3d energy-momentum tensor $T^{(3)}_{\mu\nu}$ will depend on the details of the 3d theory, which we do not specify. We need $T^{(3)}_{\mu\nu}$ to establish the conservation in directions tangent to the defect but not in order to compute the displacement as $T_{\mu n} = T^{(4)}_{\mu n}$ from \eqref{Tmn_comb_app}. This leads to
\beq
\label{gague-dis}
\partial^\mu T_{\mu n} = \delta(x^n)F_{n k} j_{(3)}^k.
\eeq

\section{Superspace conventions and useful formulas}
\label{app:conv}

\subsection{4d superspace} \label{4d_superspace_app}
Our conventions follow quite closely Wess and Bagger. For convenience we mention here a few formulas which are used in the paper. The superspace coordinates are $(x^\mu, \theta, \b \theta)$ and the chiral combination is $y^{\mu } = x^\mu + i\theta\sigma^\mu \b \theta$. A chiral superfield is a function of $(y^\mu,\theta)$
\begin{align} \label{}
\Phi(y^\mu, \theta) = \phi + \sqrt2 \theta \psi + \theta^2 F. 
\end{align}
On several occasions we use a general real multiplet given by the following $\theta$ expansion
\bal \label{real_mult_app}
V &= C + i \theta \chi - i \b \theta \b \chi + \frac{i}{2} \theta^2 M - \frac{i}{2} \b \theta^2 \b M - \theta \sigma^\mu \b \theta v_\mu \cr
&\quad + i \theta^2 \b \theta\left( \b \lambda + \frac{i}{2} \b\sigma^\mu \d_\mu \chi\right) - i \b \theta^2 \theta\left( \lambda + \frac{i}{2} \sigma^\mu\d_\mu \b \chi\right) + \frac{1}{2} \theta^2 \b \theta^2 \left( D+ \frac{1}{2}\d^2 C\right).
\eal
In fact, it will be much more convenient for us to define the component fields by taking bottom component of $V$ acted upon by covariant derivative. That is
\bal \label{V_comp_app}
&V| = C, & & D_\alpha V | = i \chi_{\alpha}, && \b D_{\dot\alpha} V| = - i \b \chi_{\dot\alpha }, \cr
& D^2 V| = -2iM, && \b D^2 V| = 2i \b M, && [D_\alpha,\b D_{\dot\alpha} ] V| = -2v_{\alpha\dot\alpha}, \cr
& \b D^2 D_\alpha V| = 4i \lambda_{\alpha},\quad && D^2 \b D_{\dot\alpha} V| = -4 i {\b\lambda}_{\dot\alpha}, \quad&&
D^\alpha \b D^2 D_{\alpha} V| =8 D.
\eal
Also useful:
\begin{align} \label{}
D_\alpha \b D_{\dot\alpha} V| = - i \d_{\alpha\dot\alpha}C - v_{\alpha\dot\alpha}, &&
\b D_{\dot\alpha} D_\alpha V| = - i \d_{\alpha\dot\alpha}C + v_{\alpha\dot\alpha}.
\end{align}
To analyse the energy-momentum multiplets we also consider a vector real multiplet $V_\mu = C_\mu + i \theta \chi_\mu + \cdots$. All the formulas above are applied by adding a vector index in an obvious way. For example $[D_\alpha,\b D_{\dot\alpha} ] V_\mu| = -2v_{\alpha\dot\alpha\mu}$. The following covariant derivatives identities are useful: 
\begin{align} \label{}
&[\b D_{\dot\alpha}, D^2] = 4i D^\alpha \d_{\alpha\dot\alpha}, 
&& [D^{\alpha}, \b D^2] = 4i \b D_{\dot\alpha}\d^{\dot\alpha\alpha}.
\end{align}
This form is far superior than the $\theta$ expansion in terms of the efficiency of computations. 

We use two other chiral superfields which are derived from $V$. We write them here as reference. 
The first, $\b D^2 V$, is in components (in $(y,\theta)$ coordinates)
\begin{align} \label{chiral_from_V}
\b D^2 V &= 2i \b M + 4i \theta(\lambda+ i\sigma^\mu \d_\mu \b \chi) -2 \theta^2(D+ \d^2 C- i \d_\mu v^\mu).
\end{align}
This arises in the context of the current multiplet, which is a real multiplet satisfying $\b D^2 V = 0$. 
The second chiral superfield is the field strength associated with $V$ viewed as an Abelian gauge multiplet
\begin{align} \label{V_field_strength}
W_\alpha &= -\frac{1}{4} \b D^2 D_\alpha V
= - i \lambda_\alpha + \theta_\alpha D - i (\sigma^{\mu\nu}\theta)_\alpha F_{\mu\nu} + \theta^2 (\sigma^\mu \d_\mu \b \lambda)_\alpha. 
\end{align}

\subsection{3d superspace} \label{3d_superspace_app}

The 3d superspace has coordinates $(x'^{i}, \Theta'_\alpha)$. To embed it in the 4d superspace we define new fermionic coordinates 
\begin{align} \label{}
\Theta_\alpha = \frac{1}{\sqrt{2}} ( \theta+ \sigma^n \b \theta)_\alpha, \qquad
\t \Theta_\alpha = \frac{i}{\sqrt{2}} ( \theta - \sigma^n \b \theta)_\alpha,
\end{align}
and $\tilde x^n = x^n - \frac{i}{2}(\theta^2 - \b\theta^2)$. The embedding is given by $(x^{i}, \tilde x^n , \Theta_\alpha, \t\Theta_\alpha) = (x'^{i}, 0, \Theta'_\alpha, 0)$. In practice we identify $x^{i} = x'^{i}$, $\Theta_\alpha = \Theta'_\alpha$ and forget about the tilded coordinates. We have also described in the paper an embedding in the chiral superspace $(y^{
\mu}, \theta_\alpha)$ which is similarly defined. As explained, the motivation for this definition is that the subspace is invariant under the super-algebra preserved by the defect. 
The following relations are easily derived
\begin{gather} \label{}
\theta^2 = \frac{1}{2}(\Theta^2 - \t\Theta^2) - i \Theta \t\Theta, \qquad
\b \theta^2 = \frac{1}{2}(\Theta^2 - \t\Theta^2) + i \Theta \t\Theta, \\
\theta \sigma^\mu \b \theta =  \frac{1}{2}(\Theta^2 + \t\Theta^2), \qquad
\theta \sigma^i \b \theta = i \Theta \Gamma^i \t\Theta, \qquad
\theta^2 \b \theta^2 = - \Theta^2 \t \Theta^2.
\end{gather}
The change of basis in the 4d superspace is accompanied with the associated covariant derivatives
\begin{align} \label{}
\Delta_\alpha &= \frac{1}{\sqrt{2}} \left(D_\alpha + (\sigma^n \b D)_\alpha \right) 
= \frac{\d}{\d \Theta^\alpha} + i(\Gamma^{i}\Theta)_\alpha \d_{i} - \t \Theta_\alpha \d_n, \\
\t \Delta_\alpha &= -\frac{i}{\sqrt{2}} \left(D_\alpha - (\sigma^n \b D)_\alpha \right)
= \frac{\d}{\d \t \Theta^\alpha} + i(\Gamma^{i}\t \Theta)_\alpha \d_{i} + \Theta_\alpha \d_n,
\end{align}
which satisfy 
\bal
&\{ \Delta_\alpha, \Delta^\beta \} = -2i (\Gamma^{i})_\alpha{}^\beta \d_{i}, \qquad
\Delta^\alpha \Delta_\beta \Delta_\alpha = 0, \qquad
\Delta^2 \Delta^2 = 4 \d_i \d^i, \cr
&\Delta_\alpha \Delta_\beta = -i\d_{\alpha\beta} + \tfrac{1}{2}\epsilon_{\alpha\beta} \Delta^2, \qquad
\Delta^2 \Delta_\alpha = - \Delta_\alpha \Delta^2 = -2 i \d_{\alpha\beta} \Delta^\beta.
\eal
(Similarly for $\t \Delta$.) 
For bookkeeping, we present the following relations for converting covariant derivatives in the different bases
\begin{align} 
\Delta^2 &= \frac{1}{2}(D^2 + \b D^2) + D\sigma^n \b D - 2i \d_n \label{Delta_to_D1} \\
&= \frac{1}{2}(D^2 + \b D^2) - \b D\b\sigma^n D + 2i \d_n, \label{Delta_to_D2} \\
i\Delta^\alpha \t \Delta_\alpha &= \frac{1}{2}(D^2 - \b D^2) - 2i \d_n, \label{Delta_covder_relation} \\
\Delta^{(\beta} \t \Delta_{\alpha)} &= \frac{i}{2} \left( (\b D\b \sigma^n)^{(\beta} D_{\alpha)} - (\sigma^n \b D)_{(\alpha} D^{\beta)} \right), \\
\sqrt2 i \Delta^2 \t \Delta_\alpha &= \b D^2 D_\alpha - D^2 (\sigma^n \b D)_\alpha + 2i \Gamma^j (D- \sigma^n \b D)_\alpha \d_j. 
\end{align}
These are useful for computing the 3d components of 4d superfields. As an example, consider the decomposition of the 4d real multiplet $V$ \eqref{real_mult_app}. We find
\begin{align} \label{}
V|_{\t\Theta=0} &= C + \frac{i}{\sqrt2} \Theta (\chi-\sigma^n \b \chi) + \frac{1}{2} \Theta^2 \left( \frac{i}{2}(M- \b M) -v_n\right), \\
\t\Delta_\alpha V|_{\t\Theta=0} &= \frac{1}{\sqrt2}(\chi + \sigma^n \b \chi)_\alpha + \Theta_\alpha \left(\frac{1}{2}(M+\b M) + \d_n C\right) 
+ i (\Gamma^j \Theta)_\alpha v_j  \cr
&\quad - \frac{1}{2} \Theta^2 \left( \sqrt2 (\lambda+ \sigma^n \b \lambda)_\alpha + \frac{i}{\sqrt2} \big(\Gamma^j \d_j(\chi + \sigma^n \b \chi)\big)_\alpha\right).
\nonumber\end{align}
$\t\Delta^2 V|_{\t\Theta=0}$ can be computed similarly but we shall not need it. To demonstrate this computation let consider the $\Theta_\alpha$ component of $\t\Delta_\alpha V|_{\t\Theta=0}$. It is obtained by applying the covariant derivative and using \eqref{Delta_covder_relation}. This leads to
\begin{align} \label{}
-\frac{1}{2} \Delta^\alpha \t \Delta_\alpha V| = \frac{i}{4} (D^2 - \b D^2)V| +  \d_n V|,
\end{align}
which together with \eqref{V_comp_app} can be expressed in terms of the components of $V$.

\section{The $\cS$-multiplet} \label{s_mult_app}
\label{app:S}

In our conventions the $\cS$-multiplet \cite{Komargodski:2010rb} $\cS_{\alpha\dot\alpha} = \sigma^\mu_{\alpha\dot\alpha} \cS_\mu$ is given by 
\begin{align} \label{}
\b D^{\dot\alpha} \cS_{\alpha\dot\alpha} = 2(\chi_\alpha - \cY_\alpha). 
\end{align}
Here $\chi_\alpha$ satisfies $D^\alpha \chi_\alpha = \b D_{\dot\alpha} \b \chi^{\dot\alpha}$. In components this is solved by
\begin{align} \label{}
\chi_\alpha = -i \lambda_\alpha + \theta_\alpha D - i (\sigma^{\mu\nu}\theta)_\alpha F_{\mu\nu} + \theta^2 (\sigma^{\mu} \d_\mu \b \lambda)_\alpha,
\end{align}
with $D$ real and $F_{\mu\nu} = - F_{\nu\mu}$ satisfying the Bianchi identity, that is it can locally be written as $F_{\mu\nu} = \d_\mu v_\nu - \d_\nu v_\mu$. In addition, we can locally define a chiral superfield $X=x+ \sqrt2 \theta\psi + \theta^2 F$ such that $\cY_\alpha = D_\alpha X$. Solving for the components of $\cS_\mu$ gives
\bal \label{S_mult_exp_app}
\cS_\mu &= j_\mu - i\theta \left(S_\mu - 2\sqrt2 i \sigma_\mu \b \psi \right) + i\b\theta \left( \b S_\mu -2 \sqrt2 i \b \sigma_\mu \psi\right) + 2 i\theta^2 \d_\mu \b x - 2i \b \theta^2 \d_\mu x \cr
&\quad + \theta \sigma^\nu \b \theta \left( 2T_{\nu\mu} - 4\eta_{\nu\mu} A - \frac{1}{2} \epsilon_{\nu\mu\rho\sigma}\left( \d^\rho j^\sigma - F^{\rho\sigma}\right) \right) \cr
&\quad - \frac{1}{2} \theta^2 \b \theta\left(\b \sigma^\nu \d_\nu S_\mu + 2\sqrt2 i\b \sigma_\mu \sigma^\nu \d_\nu \b\psi\right) + \frac{1}{2}\b\theta^2 \theta\left( \sigma^\nu \d_\nu \b S_{\mu} + 2\sqrt 2i \sigma_\mu \b \sigma^\nu \d_\nu \psi\right) \cr
&\quad + \frac{1}{2} \theta^2\b\theta^2 \left( \d_\mu \d_\nu j^\nu - \frac{1}{2} \d^2 j_\mu \right).
\eal
In this expression $S_{\alpha\mu}$ is conserved, $T_{\mu\nu}$ is symmetric and conserved, and
\begin{align} \label{}
T^\mu{}_\mu = 6 A + D, \qquad
(\sigma^\mu \b S_\mu)_\alpha = -2 \lambda_\alpha - 6 \sqrt2 i \psi_\alpha, \qquad
\d^\mu j_\mu = 4B,
\end{align}
where $F= A+ i B$.

Improvements by a real multiplet $U$ take the form
\bal
&S_{\alpha\dot\alpha} \rightarrow S_{\alpha\dot\alpha} -[D_\alpha, \b D_{\dot\alpha}]U, \\
& \chi_\alpha \rightarrow \chi_\alpha - \tfrac{3}{4}\b D^2 D_\alpha U, \\
&\cY_\alpha \rightarrow \cY_\alpha + \tfrac{1}{4} D_\alpha \b D^2 U.
\eal

For a sigma model with K\"{a}hler potential $K(\b \Phi^{\b a}, \Phi^a)$ and superpotential $W(\Phi^a)$ the $\cS$-multiplet is given by
\bal \label{S_for_kahler}
&\cS_{\alpha \dot \alpha} = K_{a \b a} \b D_{\dot \alpha} \b \Phi^{\b a} D_\alpha \Phi^a,  \\
&\chi_\alpha = -\tfrac{1}{4}\b D^2 D_\alpha K,  \\
&\cY_\alpha = D_\alpha W.
\eal
The FZ-multiplet exists if the improvement $U_{\text{FZ}} = - \frac{1}{3}K$ is well defined in which case $\chi_\alpha = 0$ and
\bal \label{FZ_for_kahler}
&\cJ_{\alpha\dot\alpha} = K_{a \b a} \b D_{\dot \alpha} \b \Phi^{\b a} D_\alpha \Phi^a + \tfrac{1}{3}[D_\alpha, \b D_{\dot\alpha} ]K, \cr
&\cY_\alpha = D_\alpha W - \tfrac{1}{12}D_\alpha \b D^2 K.
\eal
If there is an $R$-symmetry, with $R[\Phi^a]=R_a$, we may define $U_\cR = \frac{1}{2} \sum R_a \Phi^a K_a$. Using the equations of motion $\b D^2 K_n = 4 W_n$ this leads to $\cY_\alpha =0$ and
\bal \label{R_for_kahler}
&\cR_{\alpha\dot\alpha} = K_{a \b a} \b D_{\dot \alpha} \b \Phi^{\b a} D_\alpha \Phi^a - [D_\alpha, \b D_{\dot\alpha} ]U_\cR, \cr
&\chi_\alpha = -\tfrac{1}{4}\b D^2 D_\alpha (K + 3 U_\cR).
\eal

\bibliographystyle{JHEP}
\bibliography{disp}

\providecommand{\href}[2]{#2}\begingroup\raggedright\begin{thebibliography}{10}

\bibitem{Polyakov:2000ti}
A.~M. Polyakov and V.~S. Rychkov, \emph{{Gauge field strings duality and the
  loop equation}},
  \href{http://dx.doi.org/10.1016/S0550-3213(00)00183-8}{\emph{Nucl. Phys.}
  {\bf B581} (2000) 116--134},
  [\href{https://arxiv.org/abs/hep-th/0002106}{{\tt hep-th/0002106}}].

\bibitem{Semenoff:2004qr}
G.~W. Semenoff and D.~Young, \emph{{Wavy Wilson line and $AdS$/CFT}},
  \href{http://dx.doi.org/10.1142/S0217751X0502077X}{\emph{Int. J. Mod. Phys.}
  {\bf A20} (2005) 2833--2846},
  [\href{https://arxiv.org/abs/hep-th/0405288}{{\tt hep-th/0405288}}].

\bibitem{Correa:2012at}
D.~Correa, J.~Henn, J.~Maldacena and A.~Sever, \emph{{An exact formula for the
  radiation of a moving quark in ${\cal N}=4$ super Yang Mills}},
  \href{http://dx.doi.org/10.1007/JHEP06(2012)048}{\emph{JHEP} {\bf 06} (2012)
  048}, [\href{https://arxiv.org/abs/arXiv:1202.4455}{{\tt arXiv:1202.4455}}].

\bibitem{Bianchi:2015liz}
L.~Bianchi, M.~Meineri, R.~C. Myers and M.~Smolkin, \emph{{R\'enyi entropy and
  conformal defects}},
  \href{http://dx.doi.org/10.1007/JHEP07(2016)076}{\emph{JHEP} {\bf 07} (2016)
  076}, [\href{https://arxiv.org/abs/arXiv:1511.06713}{{\tt
  arXiv:1511.06713}}].

\bibitem{Dong:2016wcf}
X.~Dong, \emph{{Shape dependence of holographic R\'enyi entropy in conformal
  field theories}},
  \href{http://dx.doi.org/10.1103/PhysRevLett.116.251602}{\emph{Phys. Rev.
  Lett.} {\bf 116} (2016) 251602},
  [\href{https://arxiv.org/abs/arXiv:1602.08493}{{\tt arXiv:1602.08493}}].

\bibitem{Balakrishnan:2016ttg}
S.~Balakrishnan, S.~Dutta and T.~Faulkner, \emph{{Gravitational dual of the
  R\'{e}nyi twist displacement operator}},
  \href{https://arxiv.org/abs/arXiv:1607.06155}{{\tt arXiv:1607.06155}}.

\bibitem{Bianchi:2016xvf}
L.~Bianchi, S.~Chapman, X.~Dong, D.~A. Galante, M.~Meineri and R.~C. Myers,
  \emph{{Shape dependence of holographic R\'enyi entropy in general
  dimensions}}, \href{http://dx.doi.org/10.1007/JHEP11(2016)180}{\emph{JHEP}
  {\bf 11} (2016) 180}, [\href{https://arxiv.org/abs/arXiv:1607.07418}{{\tt
  arXiv:1607.07418}}].

\bibitem{Chu:2016tps}
C.-S. Chu and R.-X. Miao, \emph{{Universality in the shape dependence of
  holographic R\'enyi entropy for general higher derivative gravity}},
  \href{http://dx.doi.org/10.1007/JHEP12(2016)036}{\emph{JHEP} {\bf 12} (2016)
  036}, [\href{https://arxiv.org/abs/arXiv:1608.00328}{{\tt
  arXiv:1608.00328}}].

\bibitem{Jensen:2015swa}
K.~Jensen and A.~O'Bannon, \emph{{Constraint on defect and boundary
  renormalization group flows}},
  \href{http://dx.doi.org/10.1103/PhysRevLett.116.091601}{\emph{Phys. Rev.
  Lett.} {\bf 116} (2016) 091601},
  [\href{https://arxiv.org/abs/arXiv:1509.02160}{{\tt arXiv:1509.02160}}].

\bibitem{Billo:2016cpy}
M.~Bill\`o, V.~Gon\c{c}alves, E.~Lauria and M.~Meineri, \emph{{Defects in
  conformal field theory}},
  \href{http://dx.doi.org/10.1007/JHEP04(2016)091}{\emph{JHEP} {\bf 04} (2016)
  091}, [\href{https://arxiv.org/abs/arXiv:1601.02883}{{\tt
  arXiv:1601.02883}}].

\bibitem{Gadde:2016fbj}
A.~Gadde, \emph{{Conformal constraints on defects}},
  \href{https://arxiv.org/abs/arXiv:1602.06354}{{\tt arXiv:1602.06354}}.

\bibitem{Gaiotto:2013sma}
D.~Gaiotto, S.~Gukov and N.~Seiberg, \emph{{Surface defects and resolvents}},
  \href{http://dx.doi.org/10.1007/JHEP09(2013)070}{\emph{JHEP} {\bf 09} (2013)
  070}, [\href{https://arxiv.org/abs/arXiv:1307.2578}{{\tt arXiv:1307.2578}}].

\bibitem{Komargodski:2010rb}
Z.~Komargodski and N.~Seiberg, \emph{{Comments on supercurrent multiplets,
  supersymmetric field theories and supergravity}},
  \href{http://dx.doi.org/10.1007/JHEP07(2010)017}{\emph{JHEP} {\bf 07} (2010)
  017}, [\href{https://arxiv.org/abs/arXiv:1002.2228}{{\tt arXiv:1002.2228}}].

\bibitem{Ferrara:1974pz}
S.~Ferrara and B.~Zumino, \emph{{Transformation properties of the
  supercurrent}},
  \href{http://dx.doi.org/10.1016/0550-3213(75)90063-2}{\emph{Nucl. Phys.} {\bf
  B87} (1975) 207}.

\bibitem{Gates:1983nr}
S.~J. Gates, M.~T. Grisaru, M.~Rocek and W.~Siegel, \emph{{Superspace or one
  thousand and one lessons in supersymmetry}}, {\emph{Front. Phys.} {\bf 58}
  (1983) 1--548}, [\href{https://arxiv.org/abs/hep-th/0108200}{{\tt
  hep-th/0108200}}].

\bibitem{Wess:1992cp}
J.~Wess and J.~Bagger, \emph{{Supersymmetry and supergravity}}.
\newblock 1992.

\bibitem{Kuzenko:2011xg}
S.~M. Kuzenko, U.~Lindstrom and G.~Tartaglino-Mazzucchelli, \emph{{Off-shell
  supergravity-matter couplings in three dimensions}},
  \href{http://dx.doi.org/10.1007/JHEP03(2011)120}{\emph{JHEP} {\bf 03} (2011)
  120}, [\href{https://arxiv.org/abs/arXiv:1101.4013}{{\tt arXiv:1101.4013}}].

\bibitem{Kuzenko:2010rp}
S.~M. Kuzenko, J.-H. Park, G.~Tartaglino-Mazzucchelli and R.~Unge,
  \emph{{Off-shell superconformal nonlinear sigma-models in three dimensions}},
  \href{http://dx.doi.org/10.1007/JHEP01(2011)146}{\emph{JHEP} {\bf 01} (2011)
  146}, [\href{https://arxiv.org/abs/arXiv:1011.5727}{{\tt arXiv:1011.5727}}].

\bibitem{Buchbinder:2015qsa}
E.~I. Buchbinder, S.~M. Kuzenko and I.~B. Samsonov, \emph{{Superconformal field
  theory in three dimensions: Correlation functions of conserved currents}},
  \href{http://dx.doi.org/10.1007/JHEP06(2015)138}{\emph{JHEP} {\bf 06} (2015)
  138}, [\href{https://arxiv.org/abs/arXiv:1503.04961}{{\tt
  arXiv:1503.04961}}].

\bibitem{Dumitrescu:2011iu}
T.~T. Dumitrescu and N.~Seiberg, \emph{{Supercurrents and brane currents in
  diverse dimensions}},
  \href{http://dx.doi.org/10.1007/JHEP07(2011)095}{\emph{JHEP} {\bf 07} (2011)
  095}, [\href{https://arxiv.org/abs/arXiv:1106.0031}{{\tt arXiv:1106.0031}}].

\bibitem{Bilal:2011gp}
A.~Bilal, \emph{{Supersymmetric boundaries and junctions in four dimensions}},
  \href{http://dx.doi.org/10.1007/JHEP11(2011)046}{\emph{JHEP} {\bf 11} (2011)
  046}, [\href{https://arxiv.org/abs/arXiv:1103.2280}{{\tt arXiv:1103.2280}}].

\bibitem{DeWolfe:2001pq}
O.~DeWolfe, D.~Z. Freedman and H.~Ooguri, \emph{{Holography and defect
  conformal field theories}},
  \href{http://dx.doi.org/10.1103/PhysRevD.66.025009}{\emph{Phys. Rev.} {\bf
  D66} (2002) 025009}, [\href{https://arxiv.org/abs/hep-th/0111135}{{\tt
  hep-th/0111135}}].

\bibitem{Mintun:2014aka}
E.~Mintun, J.~Polchinski and S.~Sun, \emph{{The field theory of intersecting
  D3-branes}}, \href{http://dx.doi.org/10.1007/JHEP08(2015)118}{\emph{JHEP}
  {\bf 08} (2015) 118}, [\href{https://arxiv.org/abs/arXiv:1402.6327}{{\tt
  arXiv:1402.6327}}].

\bibitem{Erdmenger:2002ex}
J.~Erdmenger, Z.~Guralnik and I.~Kirsch, \emph{{Four-dimensional superconformal
  theories with interacting boundaries or defects}},
  \href{http://dx.doi.org/10.1103/PhysRevD.66.025020}{\emph{Phys. Rev.} {\bf
  D66} (2002) 025020}, [\href{https://arxiv.org/abs/hep-th/0203020}{{\tt
  hep-th/0203020}}].

\bibitem{Constable:2002xt}
N.~R. Constable, J.~Erdmenger, Z.~Guralnik and I.~Kirsch, \emph{{Intersecting
  D3 branes and holography}},
  \href{http://dx.doi.org/10.1103/PhysRevD.68.106007}{\emph{Phys. Rev.} {\bf
  D68} (2003) 106007}, [\href{https://arxiv.org/abs/hep-th/0211222}{{\tt
  hep-th/0211222}}].

\bibitem{Osborn:1998qu}
H.~Osborn, \emph{{${\cal N}=1$ superconformal symmetry in four-dimensional
  quantum field theory}},
  \href{http://dx.doi.org/10.1006/aphy.1998.5893}{\emph{Annals Phys.} {\bf 272}
  (1999) 243--294}, [\href{https://arxiv.org/abs/hep-th/9808041}{{\tt
  hep-th/9808041}}].

\bibitem{Magro:2001aj}
M.~Magro, I.~Sachs and S.~Wolf, \emph{{Superfield Noether procedure}},
  \href{http://dx.doi.org/10.1006/aphy.2002.6239}{\emph{Annals Phys.} {\bf 298}
  (2002) 123--166}, [\href{https://arxiv.org/abs/hep-th/0110131}{{\tt
  hep-th/0110131}}].

\bibitem{Buchbinder:1998qv}
I.~L. Buchbinder and S.~M. Kuzenko, \emph{{Ideas and methods of supersymmetry
  and supergravity: Or a walk through superspace}}.
\newblock 1998.

\bibitem{Kuzenko:2010ni}
S.~M. Kuzenko, \emph{{Variant supercurrents and Noether procedure}},
  \href{http://dx.doi.org/10.1140/epjc/s10052-010-1513-1}{\emph{Eur. Phys. J.}
  {\bf C71} (2011) 1513}, [\href{https://arxiv.org/abs/arXiv:1008.1877}{{\tt
  arXiv:1008.1877}}].

\bibitem{West:1997vm}
P.~C. West, \emph{{Introduction to rigid supersymmetric theories}},  in
  \emph{{Confinement, duality, and nonperturbative aspects of QCD. Proceedings,
  NATO Advanced Study Institute, Newton Institute Workshop, Cambridge, UK, June
  23-July 4, 1997}}, pp.~87--150, 1997.
\newblock \href{https://arxiv.org/abs/hep-th/9805055}{{\tt hep-th/9805055}}.

\bibitem{Park:1999cw}
J.-H. Park, \emph{{Superconformal symmetry in three-dimensions}},
  \href{http://dx.doi.org/10.1063/1.1290056}{\emph{J. Math. Phys.} {\bf 41}
  (2000) 7129--7161}, [\href{https://arxiv.org/abs/hep-th/9910199}{{\tt
  hep-th/9910199}}].

\bibitem{Festuccia:2011ws}
G.~Festuccia and N.~Seiberg, \emph{{Rigid supersymmetric theories in curved
  superspace}}, \href{http://dx.doi.org/10.1007/JHEP06(2011)114}{\emph{JHEP}
  {\bf 06} (2011) 114}, [\href{https://arxiv.org/abs/arXiv:1105.0689}{{\tt
  arXiv:1105.0689}}].

\bibitem{Belyaev:2005rt}
D.~V. Belyaev, \emph{{Boundary conditions in supergravity on a manifold with
  boundary}},
  \href{http://dx.doi.org/10.1088/1126-6708/2006/01/047}{\emph{JHEP} {\bf 01}
  (2006) 047}, [\href{https://arxiv.org/abs/hep-th/0509172}{{\tt
  hep-th/0509172}}].

\bibitem{Belyaev:2007bg}
D.~V. Belyaev and P.~van Nieuwenhuizen, \emph{{Tensor calculus for supergravity
  on a manifold with boundary}},
  \href{http://dx.doi.org/10.1088/1126-6708/2008/02/047}{\emph{JHEP} {\bf 02}
  (2008) 047}, [\href{https://arxiv.org/abs/arXiv:0711.2272}{{\tt
  arXiv:0711.2272}}].

\bibitem{Belyaev:2008ex}
D.~V. Belyaev and P.~van Nieuwenhuizen, \emph{{Simple $d=4$ supergravity with a
  boundary}},
  \href{http://dx.doi.org/10.1088/1126-6708/2008/09/069}{\emph{JHEP} {\bf 09}
  (2008) 069}, [\href{https://arxiv.org/abs/arXiv:0806.4723}{{\tt
  arXiv:0806.4723}}].

\bibitem{Andrianopoli:2014aqa}
L.~Andrianopoli and R.~D'Auria, \emph{{${\cal N}=1$ and ${\cal N}=2$ pure
  supergravities on a manifold with boundary}},
  \href{http://dx.doi.org/10.1007/JHEP08(2014)012}{\emph{JHEP} {\bf 08} (2014)
  012}, [\href{https://arxiv.org/abs/arXiv:1405.2010}{{\tt arXiv:1405.2010}}].

\bibitem{DiPietro:2015zia}
L.~Di~Pietro, N.~Klinghoffer and I.~Shamir, \emph{{On supersymmetry, boundary
  actions and brane charges}},
  \href{http://dx.doi.org/10.1007/JHEP02(2016)163}{\emph{JHEP} {\bf 02} (2016)
  163}, [\href{https://arxiv.org/abs/arXiv:1502.05976}{{\tt
  arXiv:1502.05976}}].

\bibitem{Aprile:2016gvn}
F.~Aprile and V.~Niarchos, \emph{{$ \mathcal{N} =2$ supersymmetric field
  theories on 3-manifolds with $A$-type boundaries}},
  \href{http://dx.doi.org/10.1007/JHEP07(2016)126}{\emph{JHEP} {\bf 07} (2016)
  126}, [\href{https://arxiv.org/abs/arXiv:1604.01561}{{\tt
  arXiv:1604.01561}}].

\bibitem{Lamy-Poirier:2014sea}
J.~Lamy-Poirier, \emph{{Localization of a supersymmetric gauge theory in the
  presence of a surface defect}},
  \href{https://arxiv.org/abs/arXiv:1412.0530}{{\tt arXiv:1412.0530}}.

\bibitem{Bullimore:2014upa}
M.~Bullimore and H.-C. Kim, \emph{{The superconformal index of the $(2,0)$
  theory with defects}},
  \href{http://dx.doi.org/10.1007/JHEP05(2015)048}{\emph{JHEP} {\bf 05} (2015)
  048}, [\href{https://arxiv.org/abs/arXiv:1412.3872}{{\tt arXiv:1412.3872}}].

\bibitem{Gomis:2016ljm}
J.~Gomis, B.~Le~Floch, Y.~Pan and W.~Peelaers, \emph{{Intersecting surface
  defects and two-dimensional CFT}},
  \href{https://arxiv.org/abs/arXiv:1610.03501}{{\tt arXiv:1610.03501}}.

\bibitem{Pan:2016fbl}
Y.~Pan and W.~Peelaers, \emph{{Intersecting surface defects and instanton
  partition functions}},  \href{https://arxiv.org/abs/arXiv:1612.04839}{{\tt
  arXiv:1612.04839}}.

\bibitem{Closset:2013vra}
C.~Closset, T.~T. Dumitrescu, G.~Festuccia and Z.~Komargodski, \emph{{The
  reometry of supersymmetric partition functions}},
  \href{http://dx.doi.org/10.1007/JHEP01(2014)124}{\emph{JHEP} {\bf 01} (2014)
  124}, [\href{https://arxiv.org/abs/arXiv:1309.5876}{{\tt arXiv:1309.5876}}].

\bibitem{Closset:2014uda}
C.~Closset, T.~T. Dumitrescu, G.~Festuccia and Z.~Komargodski, \emph{{From
  rigid supersymmetry to twisted holomorphic theories}},
  \href{http://dx.doi.org/10.1103/PhysRevD.90.085006}{\emph{Phys. Rev.} {\bf
  D90} (2014) 085006}, [\href{https://arxiv.org/abs/arXiv:1407.2598}{{\tt
  arXiv:1407.2598}}].

\bibitem{David:2016onq}
J.~R. David, E.~Gava, R.~K. Gupta and K.~Narain, \emph{{Localization on
  AdS$_{2} \times$ S$^{1}$}},
  \href{http://dx.doi.org/10.1007/JHEP03(2017)050}{\emph{JHEP} {\bf 03} (2017)
  050}, [\href{https://arxiv.org/abs/arXiv:1609.07443}{{\tt
  arXiv:1609.07443}}].

\bibitem{Assel:2016pgi}
B.~Assel, D.~Martelli, S.~Murthy and D.~Yokoyama, \emph{{Localization of
  supersymmetric field theories on non-compact hyperbolic three-manifolds}},
  \href{http://dx.doi.org/10.1007/JHEP03(2017)095}{\emph{JHEP} {\bf 03} (2017)
  095}, [\href{https://arxiv.org/abs/arXiv:1609.08071}{{\tt
  arXiv:1609.08071}}].

\bibitem{Genolini:2016ecx}
P.~Benetti~Genolini, D.~Cassani, D.~Martelli and J.~Sparks, \emph{{Holographic
  renormalization and supersymmetry}},
  \href{http://dx.doi.org/10.1007/JHEP02(2017)132}{\emph{JHEP} {\bf 02} (2017)
  132}, [\href{https://arxiv.org/abs/arXiv:1612.06761}{{\tt
  arXiv:1612.06761}}].

\bibitem{carlo}
P.~Liendo and C.~Meneghelli, \emph{{Bootstrap equations for $ \mathcal{N} $ = 4
  SYM with defects}},
  \href{http://dx.doi.org/10.1007/JHEP01(2017)122}{\emph{JHEP} {\bf 01} (2017)
  122}, [\href{https://arxiv.org/abs/arXiv:1608.05126}{{\tt
  arXiv:1608.05126}}].

\end{thebibliography}\endgroup

\end{document}